\documentclass{article}

\usepackage{arxiv}
\usepackage[square,numbers,sort&compress]{natbib}

\usepackage[utf8]{inputenc} 
\usepackage[T1]{fontenc}    
\usepackage{hyperref}       
\usepackage{url}            
\usepackage{booktabs}       
\usepackage{amsfonts}       
\usepackage{nicefrac}       
\usepackage{microtype}      
\usepackage{lipsum}		
\usepackage{graphicx}
\usepackage{doi}
\usepackage{tabularx}
\usepackage{graphicx}
\usepackage{subcaption}
\usepackage{amsmath}
\usepackage{mdframed}
\usepackage{longtable}
\newcolumntype{P}[1]{>{\raggedright\arraybackslash}p{#1}}
\usepackage{comment}
\usepackage{csquotes}
\usepackage[table]{xcolor}
\usepackage{multirow}
\usepackage{array}
\usepackage{placeins}  

\DeclareUnicodeCharacter{03BA}{\kappa}
\DeclareUnicodeCharacter{03BC}{\mu}
\DeclareUnicodeCharacter{03B3}{$\gamma$}
\DeclareUnicodeCharacter{03B4}{$\delta$}

\usepackage{booktabs}  
\usepackage{siunitx}   
\usepackage{graphicx}  

\usepackage{listings}
\usepackage{soul}
\usepackage{xcolor}

\usepackage{fancyvrb}

\title{The Lyme Disease Controversy:  An AI-Driven Discourse Analysis of a Quarter Century of Academic Debate and Divides}

\author{
    \href{https://orcid.org/0000-0001-9416-1435}{\includegraphics[scale=0.06]{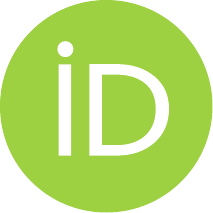}\hspace{1mm}Teo Susnjak} \thanks{Corresponding author: t.susnjak@massey.ac.nz},  
        \href{https://orcid.org/0009-0007-7499-3892}{\includegraphics[scale=0.06]{orcid.pdf}\hspace{1mm}Cole Palffy},    
    \href{https://orcid.org/0009-0008-3141-443X}{\includegraphics[scale=0.06]{orcid.pdf}\hspace{1mm}Tatiana Zimina},
    \href{https://orcid.org/0009-0002-1393-8682}{\includegraphics[scale=0.06]{orcid.pdf}\hspace{1mm}Nazgul Altynbekova}\\ 
    School of Mathematical and Computational Sciences, Massey University, Albany, New Zealand \\
    \AND
    \href{https://orcid.org/0000-0003-4346-027X}{\includegraphics[scale=0.06]{orcid.pdf}\hspace{1mm}Kunal Garg},  
    \href{https://orcid.org/0000-0002-7470-5770}{\includegraphics[scale=0.06]{orcid.pdf}\hspace{1mm}Leona Gilbert} \\  
    Te?ted Oy, Jyväskylä, Finland
}

\renewcommand{\shorttitle}{AI-Driven Discourse Analysis of the Lyme Disease Controversy}

\begin{document}
\maketitle

\begin{abstract}

The scientific discourse surrounding Chronic Lyme Disease (CLD) and Post-Treatment Lyme Disease Syndrome (PTLDS) has evolved over the past twenty-five years into a complex and polarised debate, shaped by shifting research priorities, institutional influences, and competing explanatory models. This study presents the first large-scale, systematic examination of this discourse using an innovative hybrid AI-driven methodology, combining large language models with structured human validation to analyse thousands of scholarly abstracts spanning 25 years. By integrating computational techniques with expert oversight, we developed a quantitative framework for tracking epistemic shifts in contested medical fields, with applications to other content analysis domains.
Our analysis revealed a progressive transition from infection-based models of Lyme disease to immune-mediated explanations for persistent symptoms, a shift that has been particularly pronounced in high-impact clinical and immunology journals. At the same time, research supporting CLD has remained largely confined to hypothesis-driven publications, indicating a persistent asymmetry in how competing perspectives are disseminated and legitimised. The investigation into thematic trends further highlighted the enduring complexity of Lyme disease diagnostics and evolving research focus on therapeutic controversies, even as institutional alignment with PTLDS perspectives continues to grow.
This study offers new empirical insights into the structural and epistemic forces shaping Lyme disease research, providing a scalable and replicable methodology for analysing discourse. The findings have implications for policymakers, clinicians, and communication strategists, emphasising the need for more equitable research funding, standardised diagnostic criteria, and improved patient-centred care models. This research also underscores the value of AI-assisted methodologies in social science and medical research by systematically quantifying discourse evolution, offering a foundation for future studies examining other contested conditions and controversies.

\end{abstract}

\keywords{Lyme disease controversy \and Chronic Lyme Disease (CLD) \and Post-Treatment Lyme Disease Syndrome (PTLDS) \and Medical controversy \and AI in medical research \and Large Language Models in academic analysis \and Stance detection in medical literature \and Lyme disease academic discourse \and Science and Technology Studies \and Social construction of knowledge}

\section{Introduction}

It has been estimated that every year, Lyme disease affects hundreds of thousands in North America and Europe \cite{kugeler2021estimating,sykes2017estimate}, while for over a quarter of a century, the medical and scientific communities have been sharply divided over the persistent effects of Lyme disease on patients after standard antibiotic treatments \cite{shadick1994long,maloney2016controversies,tonks2007lymewars,sigal2002lyme,kullberg2011challenge,peretti2019lyme,coiffier2021lyme}.  Although most patients recover fully, approximately 25\% of patients \cite{wormser2006clinical,marques2008chronic,aucott2013post,rebman2017clinical,ursinus2021prevalence} continue to experience symptoms like fatigue, pain, and cognitive difficulties, sparking a debate about the nature of these persistent and debilitating health issues  \cite{rebman2017living, mac2020long,kaplan2003cognitive,rebman2021symptom,shor2019chronic}. This discourse has polarised into two major schools of thought: one asserts that these symptoms are the result of a post-infectious syndrome that does not involve ongoing bacterial infection \cite{lantos2011chronic,schmid2021posttreatment}, while the other posits that a subset of Lyme disease cases may become chronic \cite{shadick1994long,maloney2016controversies,tonks2007lymewars,sigal2002lyme}, and might require prolonged antibiotic treatment due to signs of a persistent infection \cite{wormser2006clinical,steere1994treatment}. Research indicates that how Lyme disease is perceived is influenced by power dynamics in healthcare, patient advocacy, and media discussions, highlighting conflicts between medical experts and public knowledge   \cite{conrad2010social,aronowitz2001symptoms,dumit2006illnesses}.

These conflicting viewpoints have resulted in a substantial but contentious body of academic work, with researchers, healthcare providers, and patient advocacy groups frequently taking opposing positions   \cite{elliott2021value,auwaerter2011antiscience,feder2007critical,johnson2014severity}. Mainstream medical bodies, like the Infectious Diseases Society of America (IDSA), argue that post-treatment symptoms experienced by a subset of patients after completing standard antibiotic therapy can be attributed to what they define as Post-Treatment Lyme Disease Syndrome (PTLDS) \cite{wormser2006clinical}. The IDSA suggests that symptoms like fatigue and cognitive issues are probably due to immune responses or tissue damage, not ongoing infection   \cite{tonks2007lymewars}. Conversely, organisations like the International Lyme and Associated Diseases Society (ILADS) advocate for recognising chronic Lyme disease (CLD), contending that ongoing infection may be responsible for these symptoms \cite{auwaerter2011scientific}. ILADS recommends extended antibiotic regimens, pointing to contested evidence of patient improvement \cite{cameron2014evidence}. While deeply rooted in scientific inquiry, this debate has also been shaped by patient experiences, public advocacy, and extensive media attention, further complicating efforts to reach a consensus \cite{yiannakoulias2017celebrity,arney2021chronic,auwaerter2011antiscience,sood2011controversies}. To that end, this scientific controversy is also intricately linked to sociopolitical and economic influences, encompassing insurance reimbursement systems, the regulation of alternative medicine, and the stigmatisation of disputed illnesses \cite{jutel2024putting,clarke2010bmedicalization}.

At the heart of it, the Lyme disease controversy exemplifies the sociology of medical knowledge, where grassroots patient movements contest biological authority, promoting alternative diagnosis and treatment approaches \cite{epstein1996impure,rabeharisoa2004patients}. Internet and social media have significantly shaped the narrative around Lyme disease \cite{shapiro2017false,archer2018online,zeng2018sculpture,shapiro2014lyme}. Patients who feel unheard by traditional medicine have discovered online communities to share experiences and explore alternative treatment options. 
These patients, who perceive conventional healthcare as dismissive, have increasingly sought refuge in online groups, where accounts of medical neglect are validated, and alternative illness models proliferate \cite{barker2011listening, blume2017immunization}. These digital platforms serve as counter-publics that challenge prevailing scientific narratives, exemplifying what researchers in Science and Technology Studies (STS) call \enquote{epistemic resistance} to dominant biomedical paradigms \cite{hess2016undone,callon2011acting}. The proliferation of conflicting medical assertions inside online Lyme disease forums has underscored the influence of digital platforms on health beliefs and patient choices \cite{briggle2012ethics,venturini2012building}. 
These platforms have amplified the voices of those advocating for chronic Lyme disease \cite{pascal2020emergence}. Still, according to opposing voices, they have also facilitated the spread of misinformation, further complicating and sharpening the discourse \cite{pascal2020emergence}. As a result, the controversy surrounding Lyme disease has extended beyond medical journals into mainstream media \cite{shapiro2017false,zeng2018sculpture}, shaping public perception and influencing policy decisions. 

\begin{figure}[htb]
    \centering
        \centering
        \includegraphics[width=\linewidth]{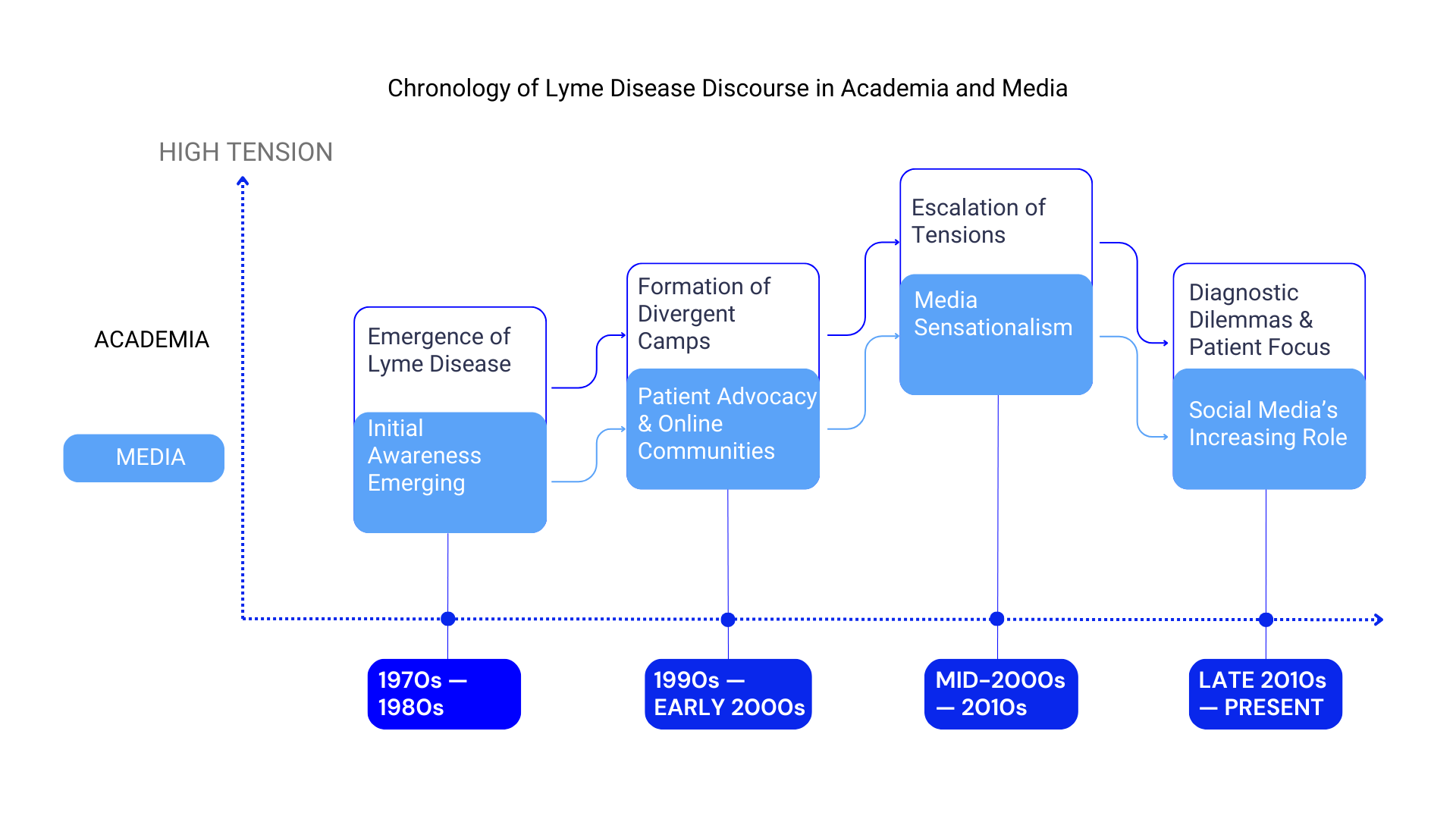}
        \caption{ A broad and consolidated outline of the discourse themes and tensions on the Lyme disease controversy in media and academia over time as reported in literature \cite{elliott2021value,stricker2008chronic,walker2021portrayal,lantos2011chronic}.}
        \label{fig:timeline}
\end{figure}

To appreciate the current state of the discourse surrounding this controversy, it is useful to chart its history. Figure \ref{fig:timeline} captures the main themes over the decades. 
During the mid-to-late 1970s, an atypical outbreak of arthritis in youngsters from rural Connecticut resulted in the early identification of what would later be termed Lyme disease \cite{steere1977epidemic}.  Shortly after its formal diagnosis, clinicians noted that several patients exhibited persistent symptoms—such as arthralgia, tiredness, and neurological complications—despite receiving prescribed antibiotic treatment \cite{steere2004emergence}. Medical researchers determined that 
Lyme disease is an illness triggered by the bacterium \textit{Borrelia burgdorferi}, transmitted through the bites of infected ticks, which acquire it from animals like mice and deer they feed on \cite{radolf2021lyme}. A hallmark of the disease in humans is a bullseye rash known as erythema migrans. Without treatment, the disease can lead to a variety of symptoms, such as joint pain and swelling, meningitis, partial facial paralysis, cognitive impairments, fatigue, headaches, and heart complications \cite{aucott2013development, stanek2011lyme}.  

In the subsequent period from the 1990s to the early 2000s, differing scientific perspectives emerged and led to opposing factions about disease chronicity and treatment, coinciding with the rise of patient advocacy and online communities that enriched public discourse. During the period from the mid-2000s to the 2010s, tensions escalated as researchers, clinicians, and patient organisations became increasingly polarised, while media coverage intensified sensationalism surrounding the chronic Lyme disease controversy  \cite{elliott2021value,stricker2008chronic,walker2021portrayal,lantos2011chronic}. 
At this time, patient organisations began to assert that the treatment protocols established by the IDSA had been employed to refuse medical insurance for those seeking extended antibiotic courses. At the same time, it also penalised physicians who prescribed these treatments \cite{elliott2021value}. 
The discourse became inflammatory with terms like \enquote{Axis of Evil} used to describe physicians prescribing prolonged antibiotics, \enquote{specialty laboratories} offering alternative tests, and the internet's role in promoting \enquote{Lyme hysteria} \cite{stricker2008chronic}, reflecting the intensifying acrimony and the frustration within the mainstream medical community. 
While it has been suggested \cite{walker2021portrayal} that Lyme disease was receiving an over-proportionate exposure in media coverage, the claim was that the coverage often perpetuated this perspective, portraying those advocating for CLD as misguided or promoting unscientific practices.

The cooling in tensions began to emerge only from the late 2010s, when the focus shifted to diagnostic problems and patient-centred perspectives, with social media playing a substantial role in shaping public understanding, activism, and patient support \cite{brunton2017stakeholder}.
As the debate continued, there has been a further shift in focus in recent years, emphasising the complexities of Lyme disease diagnostics and patients' subjective experiences. Studies now point out the challenges with current blood tests, especially in the advanced stages of the disease \cite{radolf2021lyme}. Researchers acknowledge the need for improved diagnostic tools to address potential underdiagnosis, particularly in cases with atypical symptoms or less common \textit{Borrelia burgdorferi} species~\cite{bobe2021recent}. 
Qualitative studies using interviews and ethnographic methods offer valuable insights into the  experiences of individuals with persistent symptoms attributed to Lyme disease~\cite{rebman2017living}. These studies explore the challenges of navigating a complex medical system while facing scepticism and dismissal from some healthcare providers.
What began as a relatively straightforward debate over diagnosis and treatment has expanded into a complex and often polarised controversy that touches on issues of medical authority, patient autonomy, and the role of the media in shaping public understanding of health\cite{van2018controversies}, deserving of a comprehensive investigation.

This persistent discourse on CLD and PTLDS is crucial to health communication, as it highlights the downstream effects of academic research on media narratives, which in turn affects public attitudes and, ultimately, behaviours regarding their health choices. 
It is equally important to consider the public's trust in scientific expertise and healthcare institutions in the face of stark disagreements, whose erosion of credibility can also result in various health behaviours, including non-compliance with standard treatments, seeking unverified or potentially harmful remedies, and withdrawing from conventional healthcare systems \cite{finset2020effective}.
Since such medical conflicts seldom exist solely within the realm of scientific discourse but are instead socially produced through the interplay of scientific institutions, policymakers, media, and the public \cite{radolf2021lyme,aucott2013development,clarke2003biomedicalization,timmermans2010gold}, 
Lyme disease is, therefore, a prime example of a contested sickness, illustrating how medical ambiguity generates competing knowledge claims and varied treatment paradigms \cite{aronowitz2001symptoms}.

This controversy, therefore, is important and extends beyond scientific inquiry, shaping public discourse and healthcare behaviours. Media narratives, spanning news reports, documentaries, and digital platforms, often amplify conflicting perspectives, influencing public attitudes toward diagnosis and treatment \cite{auwaerter2011scientific, johnson2014severity}. As a result, patients navigating uncertainty may turn to alternative sources when they perceive mainstream reporting as dismissive, potentially leading to non-adherence to conventional medical guidance or the pursuit of extended antibiotic regimens \cite{brossard2013science, brashers2001communication}. 
A critical dimension of this debate is the erosion of trust in scientific expertise and healthcare institutions. Conflicting guidelines from organizations such as IDSA and  ILADS contribute to uncertainty, influencing patient behaviours that range from scepticism toward standard treatments to reliance on unverified or experimental therapies \cite{feder2007critical, finset2020effective}. This complex interplay between medical uncertainty, media influence, and public trust underscores the need for strategic health communication approaches that integrate evidence-based insights with patient experiences.

\subsection*{Contribution and novelty}
Despite the wealth of research and debate, no comprehensive and systematic synthesis has been conducted to map the evolution of academic discourse on Lyme disease over the past 25 years~\cite{shapiro2018lyme}. The goal of this study has been to fill this gap, provide practical insights into the development of the  scientific discourse on this controversy, and identify overarching trends.
The key contribution of this study lies in examining over a thousand relevant academic studies spanning the past quarter-century. The novelty is centred around leveraging the latest advancements in AI technologies, employing the most sophisticated LLMs to date to perform stance and viewpoint detection expressed within these studies, allowing the extraction of both explicit positions and more nuanced sentiments contained in these texts surrounding the controversy over CLD and PTLDS. 
Furthermore, we developed a novel hybrid AI-driven methodology that enabled us to automate the processing of large volumes of text to map the discourse of this controversy and its evolution over time. Our work not only conducted a comprehensive analysis of how viewpoints have shifted over time and how different journals have platformed differing positions, but we also extracted major themes charting their development over time through to the present status of the debate. 
This timely integration of computational intelligence with medical research provides unprecedented insights into this long-standing controversy  and establishes a robust AI-driven methodology to address complex debates in healthcare literature at scale.

\section{ Related Works}

In recent years, research on Lyme disease has examined its clinical, epidemiological, and sociocultural dimensions. These studies have ranged from broad narrative and scoping reviews investigating overarching challenges to systematic reviews offering targeted insights into specific aspects of diagnosis, treatment, and disease mechanisms. The literature covered several key areas: treatment and management, pathology and disease mechanisms, diagnostic controversies and challenges, epidemiology and public health as well as the sociological perspective. 

\subsection{Treatment and Management}
The efficacy of antibiotic treatment for PTLDS has been a persistent issue where early research questioned the rationale for prolonged antimicrobial therapy. \citet{lantos2015chronic} conducted a systematic review that evaluated the role of chronic co-infections such as \textit{Babesia, Anaplasma}, and \textit{Bartonella}, concluding that no compelling evidence supported their role in PTLDS or CLD. This finding challenged earlier assertions that lingering symptoms were due to persistent infections requiring long-term antibiotic regimens.
Subsequent studies reinforced these findings. \citet{rebman2020post}  reviewed PTLDS from a mechanistic perspective and argued that persistent symptoms were more likely linked to immune dysfunction and neural sensitisation rather than ongoing infection, suggesting a shift from pathogen-focused treatments. \citet{dersch2024treatment}  further challenged the antibiotic paradigm, finding no statistically significant benefit of antimicrobial therapy on quality of life, cognition, or depression while reporting an increased incidence of adverse effects. 
Beyond clinical efficacy, the risks of overtreatment became more prominent in other reviews. \citet{sebastien2023diagnosis} conducted a systematic review and found that misdiagnosis of PTLDS was widespread, ranging from 80\% to 100\% of suspected cases being incorrectly classified, leading to unnecessary and sometimes harmful antibiotic treatment.  In addition,  \citet{mattingly2020clinical} evaluated the economic burden of Lyme disease and revealed substantial healthcare costs associated with misdiagnosis and overtreatment. At the same time, \citet{Hout2018Controversies} highlighted that despite the increasing global incidence of Lyme disease, pharmaceutical investment in its treatment was lacking. Additionally, the authors claimed that the evidence and international guidelines for managing CLD remained conflicting and controversial, posing challenges to public health policy and clinical practice.

\subsection{Pathology and Disease Mechanisms}
The biological mechanisms underlying PTLDS and CLD have also remained an area of significant debate over the past few decades. \citet{marques2008chronic} provided an early framework for distinguishing between different patient groups diagnosed with CLD, recognising that many individuals lacked objective evidence of active \textit{Borrelia burgdorferi} infection. This categorisation laid the groundwork for later investigations into the nature of persistent symptoms. \citet{borgermans2014relevance} expanded upon the early framework by \cite{marques2008chronic} exploring CLD as a multifaceted clinical entity.
The review by \citet{borgermans2014relevance} suggested that CLD remained poorly understood, with ongoing debates regarding its definition, diagnosis, and treatment.
\citet{mac2020long} further contributed to this discussion by conducting a systematic review that documented the long-term effects of Lyme disease, reporting that patients frequently experienced fatigue, musculoskeletal pain, and cognitive impairment, though the exact mechanisms remain uncertain. 
Beyond symptom classification, recent works have examined the complex biological interactions underlying \textit{Borrelia burgdorferi} persistence and disease progression. \citet{bamm2019lyme} provided an integrative review of \textit{Borrelia burgdorferi} biology, host-pathogen interactions, and immune evasion strategies, emphasising that the spirochete's ability to modulate host responses may contribute to prolonged symptoms even after standard treatment. Their findings reinforced earlier hypotheses that PTLDS symptoms could stem from immune dysregulation and chronic inflammation rather than ongoing infection. Focusing on the neuropsychiatric dimensions of Lyme disease, \citet{brackett2024neuropsychiatric} linked infection to increased risks of cognitive decline, anxiety, and depression. Recently, \citet{bobe2021recent} reviewed the progress and understanding of Lyme disease in the five years preceding their study and explored the role of immune dysregulation and potential autoimmune triggers, arguing that PTLDS symptoms were more likely driven by sustained inflammation rather than persistent infection.

\subsection{Diagnostic Controversies and Challenges}
A longstanding source of clinical and research debate has also surrounded the challenges in diagnosing Lyme disease. \citet{brunton2017stakeholder} systematically reviewed stakeholder perspectives and found widespread dissatisfaction with existing diagnostic tools. Their study highlighted the disconnect between clinician scepticism and patient experiences, which, according to the authors, frequently led to diagnostic uncertainty and strained doctor-patient relationships.
Studies have investigated the specific limitations of current diagnostic methods, and thus, the diagnosis of Lyme disease remains contentious, which is marked by significant challenges in clinical practice and research. Diagnostic uncertainty frequently arises from dissatisfaction with existing testing methods, highlighting marked discrepancies between clinical and patient experiences \cite{brunton2017stakeholder}. Conventional diagnostic tests, primarily the two-tiered approach of enzyme immunoassay (ELISA) followed by immunoblotting, have notable limitations, especially in early disease stages, leading to delayed treatments and misdiagnoses \cite{aguero2015lyme,marques2015laboratory}. Additionally, regional variation in \textit{Borrelia} genospecies poses substantial obstacles, as standard assays often fail to detect less common strains, further complicating diagnosis \cite{marques2015laboratory}. 
These diagnostic challenges extend into the debate surrounding PTLDS and CLD. Divergent perspectives among medical communities exacerbate the controversy, notably seen in conflicting guidelines from influential organisations such as the IDSA and the ILADS. Such conflicts can result in misdiagnoses, inappropriate therapies, and decreased trust in healthcare institutions \cite{beaman2016lyme,van2018controversies,schoen2020challenges}. Furthermore, geographical variations in disease presentation, including Lyme-like illnesses without confirmed local \textit{Borrelia} infections, add layers of complexity to accurate disease identification and management \cite{beaman2016lyme}. Studies have noted that resolving these diagnostic and therapeutic challenges necessitates more accurate biomarkers and standardised diagnostic protocols to improve early detection and patient outcomes \cite{aguero2015lyme}.

\subsection{Epidemiology and Public Health}
Other research reviews on the study of Lyme disease have focused more on the driving environmental factors, whereas \citet{stone2017brave} took an ecological perspective, identifying climate change and tick habitat expansion as key factors influencing Lyme disease incidence. \citet{dong2022global} conducted a global meta-analysis estimating that 14.5\% of the population had been exposed to \textit{Borrelia burgdorferi}, with the highest seroprevalence in Central Europe (20.7\%) and Eastern Asia (15.9\%). This study built on earlier epidemiological assessments, such as those by \citet{van2018controversies}, who also identified climate change and tick population dynamics as key factors in the disease’s increasing range.
Meanwhile, other studies \cite{mac2019economic} focused on the economic burden imposed by Lyme disease, concluding that it is significant, particularly in the US, and to that end, justifying further research efforts in disease control and management. \citet{mattingly2020clinical} also  evaluated the financial burden of Lyme disease, highlighting significant healthcare costs and productivity losses. \citet{bobe2021recent} noted in the U.S. that federal funding for Lyme disease research remained disproportionately low relative to its public health impact, with increasing reliance on private philanthropic contributions.

\subsection{Sociological Perspective}
\citet{peretti2019lyme} examined Lyme disease as a case study in medical controversy, showing how public mistrust in health authorities fuelled competing narratives about the disease’s prevalence and management and exacerbated  by ongoing conflicts between the IDSA and ILADS organizations, which has created ambiguity concerning optimal practice guidelines  \cite{feder2007critical, wormser2006clinical}.
 \citet{olechnowicz2023perceived} introduced a sociological perspective by examining how perceptions of vulnerability influence individual behaviours and trust in information sources regarding Lyme disease. Their study validated a novel vulnerability scale, demonstrating that emotional discomfort and perceived susceptibility shape engagement in preventive behaviours, highlighting the role of psychology and public trust in shaping disease management strategies.

Studies by  \citet{pascal2020emergence} and \citet{uzzell2012whose} emphasised the impact of media and public health communication on societal attitudes, portraying Lyme disease as either a disregarded epidemic or an exaggerated controversy, highlighting its depiction as a socially constructed phenomenon shaped by media narratives, patient advocacy, medical uncertainty, and institutional prejudices.
 \citet{puppo2023social} and \citet{hinds2019heterodox} brought to focus the distinction between conventional medical institutions and alternative \enquote{Lyme-literate} viewpoints, highlighting the knowledge disparity between orthodox and heterodox discourses, while \citet{baarsma2022knowing,rebman2017living} conducted concurrent research on patient experiences, highlighting issues on medical legitimacy, diagnostic ambiguity, and the psychological ramifications of disputed illness status. The increasing confluence of medical authority, patient empowerment, and online activism as examined in \cite{dumes2020divided,singer2019empowerment,ciotti2023three}, demonstrated how self-advocacy organisations challenge institutional authority while promoting scientific fragmentation.
Meanwhile, \citet{bloor2021knowledge} demonstrated the influence of scientific uncertainty and policy inconsistency on institutional decision-making and advocacy around Lyme disease. These studies jointly contributed towards perceiving that Lyme disease is not merely a medical ailment but a politicised and socially contentious illness, highlighting the broader conflicts among research, politics, and patient-centred healthcare. To that end, it has been postulated \cite{radolf2021lyme,aucott2013development,clarke2003biomedicalization,timmermans2010gold} that medical conflicts seldom exist solely within the realm of scientific discourse; instead, they are socially produced through the interplay of scientific institutions, policymakers, media, and the public.
Lyme disease is thus a case in point, illustrating how medical ambiguity generates competing knowledge claims and varied treatment paradigms \cite{aronowitz2001symptoms}.

\citet{pascal2020emergence} demonstrated how media discourse has significantly contributed to the perception of Lyme disease as a societal issue, bolstering biomedical scepticism and patient activism. \citet{puppo2023social} illustrated that Lyme-literate medical professionals had established alternative epistemic networks that challenge prevailing medical paradigms and promote unconventional treatment protocols. Moreover, internet platforms have revolutionised the discourse surrounding Lyme disease, serving as \enquote{knowledge enclaves} as characterised by \citet{brown2020gil}, where scientific credibility is reinterpreted through collective patient experiences rather than peer-reviewed research \cite{petersen2019navigating}. This corresponded with extensive sociological research about disseminating health-related misinformation in digital contexts, reinforcing health attitudes that deviate from conventional medical guidelines \cite{kata2010postmodern,venturini2012building}.

\subsection{Study Research Questions}

While previous reviews have addressed the clinical, epidemiological, and sociocultural dimensions of Lyme disease, they have not systematically synthesised how the academic discourse on CLD and  PTLDS has evolved. This study fills that gap by employing a large-scale, AI-driven approach to map thematic trends and stance distributions. It sought to identify and track shifts in scholarly perspectives, leading to the following key research questions:

\begin{itemize}
    \item (\textbf{RQ1}) How has the academic discourse on CLD and PTLDS evolved over the past 25 years regarding research volume, thematic focus, and stance distribution?

    \item (\textbf{RQ2}) How do journal specialisation and editorial focus influence the representation of perspectives on CLD and PTLDS in the peer-reviewed literature?  

    \item (\textbf{RQ3}) What are the dominant thematic structures within the Lyme disease debate, and how do they correspond to competing explanatory models and levels of scientific consensus?  

\end{itemize}

\section{Theoretical and Methodological Grounding in Science and Technology Studies (STS) Framework}

This study employed a Science and Technology Studies (STS) framework \cite{latour2013laboratory} in conjunction with computational AI models to analyse the Lyme disease controversy. STS offers a critical lens for understanding how social, cultural, and historical factors shape scientific knowledge, particularly in biomedicine and contested illnesses  \cite{jasanoff2001handbook,latour1987science,cetina1999epistemic,callon2011acting}.  For complex and contested illnesses like Lyme disease, an STS framework is particularly valuable in enabling the examination of the interplay of social dynamics, power relations, and the very construction of medical knowledge itself, while pairing this approach with the latest AI-automation and reasoning technologies for information extraction at scale. 
Key principles and concepts from STS that guided our overall approach in formulating the technical aspects of our methodology in Section \ref{methodology}, include:

\begin{itemize}
    \item \textbf{Selective Social Construction (STS Principle):}  STS highlights that social shaping varies across scientific fields. While physics is empirically constrained, biomedicine, addressing complex human systems, is more open to social interpretation. Even within biomedicine, diagnostic categories (PTLDS vs. CLD) are more socially negotiated than, for example, the molecular biology of \textit{Borrelia burgdorferi}.
    \item \textbf{Empirical Constraint and Real Phenomena (STS Tenet):}  Our STS framework acknowledges that social construction does not equate to extreme relativism or a dismissal of empirical reality. Empirical observation, clinical data, and technological applications do constrain scientific theories. In the Lyme context, STS helps us analyse how the interpretation of patient symptoms is contested while acknowledging the underlying reality of patient suffering and the biological basis of Lyme infection as legitimate areas of scientific inquiry.
    \item \textbf{Social Values and Epistemic Commitments in Contested Fields (STS Focus):}  Lyme disease controversies highlight paradigm clashes and differing epistemic commitments, not just factual disagreements \cite{kuhn1962structure}.  Competing values, clinical priorities, economic motivations and institutional affiliations influence research agendas and data interpretation, especially in areas lacking scientific consensus.
    \item \textbf{Discourse Analysis as an STS Methodology (STS Method):}  Computationally enhanced discourse and framing analysis were essential to reveal social influences in our study. By systematically analysing language, framing, and thematic patterns in Lyme disease literature, guided by an STS framework, we sought to expose the social processes that have historically shaped the scientific conversation and contributed to the enduring Lyme disease controversy.
\end{itemize}

The adoption of the STS framework ensured a theoretically grounded and methodologically rigorous analysis of the Lyme disease debate was conducted.  The STS framework informed both our technical methodological implementation, in aspects such as prompt engineering outlined in Section \ref{classification} as well as in the formulation of the thematic analysis of Section \ref{thematic}, but also the interpretation of results, allowing us to analyse the Lyme controversy as a socially constituted and knowledge-producing phenomenon, beyond purely biomedical or clinical perspectives. The integration of STS with medical sociology \cite{cockerham2017medical}, framing theory \cite{entman1993framing}, and patient experience research \cite{rebman2017living} collectively offered a rich lens and social science framework for understanding this complex and contested illness.

\begin{figure}[tbp]
    \centering
    \includegraphics[width=0.8\linewidth]{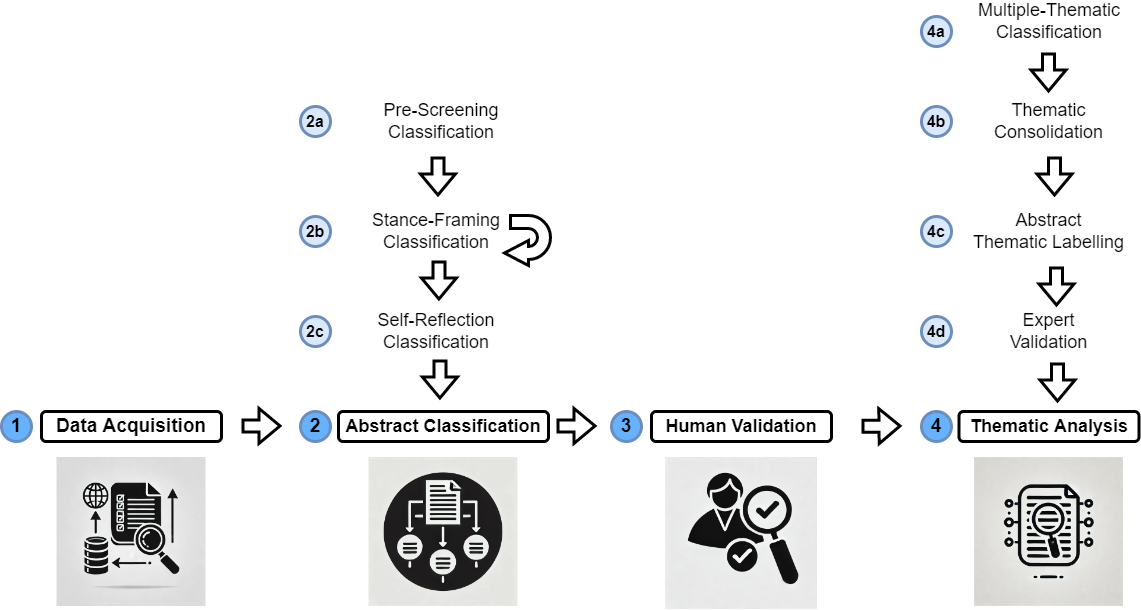}
    \caption{Overview of the steps comprising the proposed hybrid AI-driven content analysis methodology.}
    \label{fig:meth}
\end{figure}

\section{Methodology}
\label{methodology}

The technical aspects of our methodology combined the social science framework with cutting-edge AI technologies, specifically focusing on LLMs and their emerging reasoning capabilities that venture well beyond classical machine learning approaches \cite{brown2020language}. This methodological choice was motivated by the need to analyse a vast and complex body of text, exceeding the capacity of traditional qualitative methods alone. While qualitative approaches are crucial for in-depth analysis of social meaning, LLMs and their sophisticated reasoning abilities offer a scalable and systematic way to identify semantic patterns and trends across a large corpus of scientific abstracts, enabling analyses of shifts in thematic focus, stance distribution, and journal-level preferences on specific topics that would be challenging through manual review~\cite{wei2022emergent}. Increasingly, LLMs are being deployed in research for these kinds of tasks~\cite{dennstadt2024title}. Since LLMs exceed the capabilities of traditional NLP techniques, their value is found in their ability to surface implicit assumptions and underlying frames within the discourse, which can contribute  to a deeper understanding of the social construction of knowledge in this contested medical field~\cite{bender2021dangers}. We acknowledge the methodological challenges of using AI in social science research \cite{boyd2012critical}; therefore, we prioritise transparency\footnote{The datasets, LLM prompts and outputs can be found at https://github.com/teosusnjak/Lyme-disease-controversy} and validation throughout our methodology, aiming to complement and provide new approaches for scalable research rather than replace established qualitative approaches to discourse analysis. Our entire methodological approach can be summarized in four key steps that include data acquisition (Step 1), abstract classification using automated approaches (Step 2), validation of the automation process via human experts (Step 3) and finally, the thematic analysis of the dataset including the human-in-the-loop verification (Step 4). These four stages and the sub-steps are visualized in Figure \ref{fig:meth}, according to which the remainder of this section is organized.

\subsection{Dataset acquisition - Step 1}

\begin{figure}[htbp]
    \centering
    \includegraphics[width=\linewidth]{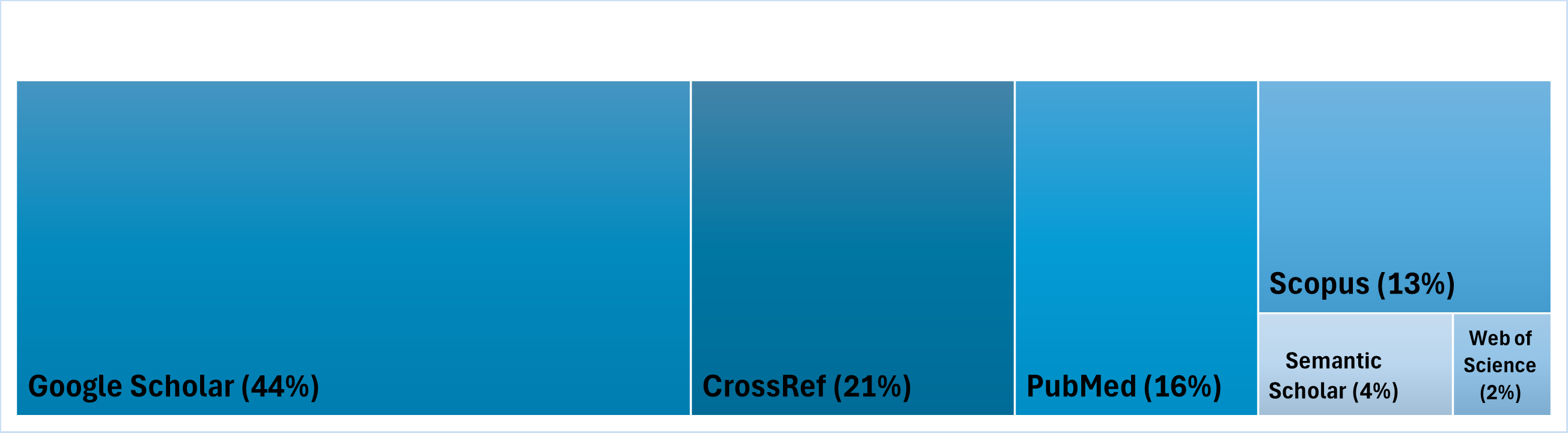}
    \caption{Proportion of retrieved articles per database}
    \label{fig:articles-by-database}
\end{figure}

Step 1 in Figure \ref{fig:meth} represents the entire data acquisition process in detail. The data was collected for a time period from 2000 to 2024 using the Publish or Perish (PoP) \cite{harzing2007pop} software for paper search~\footnote{This data collection builds on and extends the earlier work \cite{susnjak2024applying}.}. Academic databases used for paper searches were Google Scholar, Scopus, PubMed, CrossRef, Web of Science, and Semantic Scholar. Search keyword combinations were used focusing on topics related to chronic Lyme disease and post-treatment Lyme disease syndrome, and the search was restricted to each year individually due to a large number of returned results. The search term \textit{lyme} was used for paper titles. In contrast, various combinations of terms were used for matches across the paper documents comprising \textit{disease, borreliosis, borrelia, chronic, controversy, post-treatment, PTLDS, acute, syndrome, and post-Lyme}. The initial search produced a dataset of 84,140 papers covering a 25-year timeframe; however, a large proportion of abstracts were missing from this initial data collection process. Notably, Google Scholar contributed 41\% of all the retrieved records. Figure \ref{fig:articles-by-database} summarises the proportion of papers acquired by the database. The acquired dataset from PoP comprised the following fields for each paper:
\begin{itemize}
\item publication: the name of the publication source.
\item authors: the list of authors who contributed to the article.
\item year: the year the article was published, from 2000 to 2024.
\item type: paper type, i.e. article or review etc.
\item abstracts: the text of the abstract for each article - mostly unpopulated from the PoP search results. 
\item cites: the number of times that the paper has been cited. 
\end{itemize}

\subsubsection*{Data screening and filtering}

\begin{figure}
    \centering
    \includegraphics[width=1\linewidth]{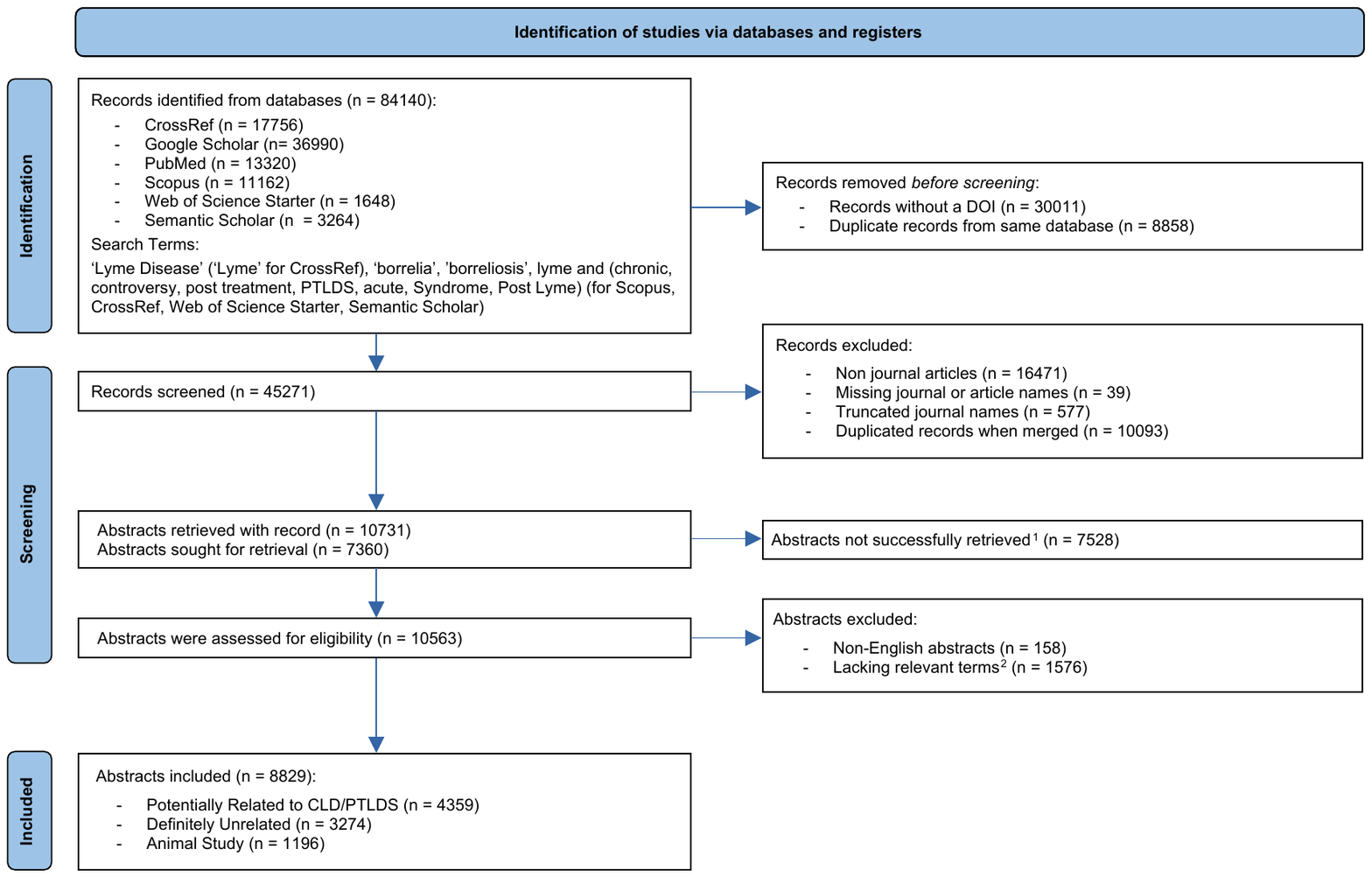}
    \caption{PRISMA flow diagram for new systematic reviews, which included searches of databases and registers only. \textsuperscript{1}This number includes 2088 abstracts with less than 300 characters in length which were rejected due to the information content being too low for analysis. \textsuperscript{2}Lyme, Borrelia*, burgdorferi, Ixodes, Erythema, migrans, tick-borne, tickborne, tick borne.}
    \label{fig:PrismaFlowDiagram}
\end{figure}

Next, the PRISMA flow diagram in Figure \ref{fig:PrismaFlowDiagram} details the pre-processing, screening and the filtering process. Records with missing or duplicated DOIs (digital object identifier) were excluded from the dataset. More than 38,000 records were removed, leaving 45,271 records for screening. Subsequently,   approximately 27,180 records were excluded due to missing publication names, article titles or abstracts. Python scripts were written to automatically retrieve missing abstracts where possible using APIs and DOIs \footnote{Scopus API  Springer Nature API } resulting in 7,360 previously missing abstracts.  Approximately 7,528 records were excluded due to irretrievable abstracts, non-English text, and lacking relevant search terms.
Consequently, the resulting dataset comprised 8829 potentially relevant abstracts, requiring more detailed screening and analysis for relevance.

\subsection{Abstract Classification - Step 2}
\label{classification}

Following the initial screening and filtering process that relied on a straightforward mechanical application of inclusion and exclusion rules, 8829 records remained, which needed a deeper semantic analysis of the content to be relevant to the PTLDs/CLD controversy. Three sub-steps with increasing sophistication were applied. Initially, an automated pre-screening classification (Step 2a) approach was applied leveraging a reasoning LLM (OpenAI's GPT-4o-mini). The purpose of this task was to eliminate the most obviously irrelevant abstracts for which a very high level of confidence in their rejection was achieved while retaining others for a deeper analysis. The aim was to remove all abstracts with a clear focus on animal studies and those that focus on Lyme disease, but hold no relevance to the PTLDS/CLD discourse. Therefore, each abstract was classified into one of three predefined classes: \textit{Potentially Related to CLD/PTLDS}, \textit{Definitely Unrelated}, or \textit{Animal Study}. This approach is supported by literature, where machine learning has been successfully used to classify medical texts for some time \cite{cohen2006reducing}. As more sophisticated models have emerged, medical text classification has improved considerably \cite{nazyrova2024leveraging}, and LLM models have thus been applied to aid in systematic reviews of the literature \cite{khraisha2023can, syriani2024screening}. Notably, due to the speed and cost, LLMs have even been found to be more accurate overall than expert annotators at classifying texts \cite{gilardi2023chatgpt}. 

Each abstract was processed individually via API calls using the classification prompt in Appendix \ref{app:prescreening}. The model generated a JSON output for every input containing the abstract’s index, a classification, and a confidence score (\textit{High}, \textit{Medium}, or \textit{Low}). Abstracts classified as \textit{Potentially Related to CLD/PTLDS} were retained for further, more refined processing and classification. Low-confidence non-\textit{Potentially Related to CLD/PTLDS} classifications was flagged for additional classification validation to ensure comprehensive coverage and minimise false exclusions. 
This low-resolution pre-screening classification step effectively categorized the 8,630 abstracts as follows: 4,160 (48.2\%) \textit{Potentially Related to CLD/PTLDS}, 3,274 (37.9\%) \textit{Definitely Unrelated}, and 1,196 (13.9\%) \textit{Animal Study}. Thus, all \textit{Animal Study} and \textit{Definitely Unrelated} abstracts receiving a Medium to High classification confidence level were eliminated from further analysis.  

\subsubsection*{Stance-Framing Classification - Step 2b}
\label{classification-first-pass}

In Step 2b, the actual classification of the remaining 4,160 abstracts into target categories that support the study's goals commenced. In this step, the authors' implicit or explicitly stated position on the controversy needed to be identified, requiring both sophisticated reasoning and domain knowledge.

\subsubsection*{From Stance to Frame Detection}

Sentiment analysis or opinion mining is often used to interpret an author's feelings on a subject~\cite{wankhade2022survey}. Frequently, the sentiment is determined by the number and strength of positive and negative words and which parts of speech are used. However, traditional sentiment analysis cannot recognise implicit meaning in text, frequently resulting in incorrect interpretations \cite{wankhade2022survey}. Sentiment analysis is ill-suited for the task of this study also because academic articles are expected to employ near-neutral expressions, while domain-specific terms such as 'disease' are neutral and objective in this context, while interpreted negatively by standard sentiment analysis models, thus skewing results for this oft appearing term.
Stance detection, on the other hand, aims to use machine learning models to automatically determine the position or attitude expressed in a text towards a specific target concept, event, or entity and seeks to identify whether the text favours, is against, or is neutral towards the target \cite{Kucuk2020stance}. Recent studies have demonstrated the effectiveness of LLMs in stance detection tasks with high reliability and accuracy \cite{lan2024stance}. 
Therefore, this research also leveraged technology for this task. 
Moreover, in Step 2b, our study conceptualised 'stance' as reflecting a more complex orientation towards the Lyme disease controversy. Drawing on framing theory \cite{entman1993framing} and discourse analysis \cite{Fairclough1992Discourse}, we defined 'stance' more broadly as the underlying perspective or interpretive \textit{frame} and underlying perspectives adopted by authors in relation to the CLD/PTLDS debate --  not merely positive/negative sentiment. 
Therefore, this included explicit agreement or disagreement with particular positions and implicit assumptions, preferred modes of reasoning, and the values and priorities emphasised in the authors' arguments. For example, a 'PTLDS-supporting stance' might be characterised by framing persistent symptoms as primarily immune-mediated, relying on epidemiological evidence, and prioritising mainstream clinical guidelines. Conversely, a CLD-supporting stance might frame persistent symptoms as indicative of ongoing infection, emphasise patient narratives and anecdotal evidence, and prioritise alternative treatment approaches. Therefore, by using LLMs for stance or \textit{frame} detection, we aim to identify these cues and more nuanced framing patterns within the academic discourse, revealing the underlying epistemological and ideological dimensions of the Lyme disease controversy. To accomplish this, we engaged with subject experts in an iterative refinement process (highlighted in Step 2b in Figure \ref{fig:meth}) to design an LLM prompt capable of performing this complex task, which also considers our STS framework. The final classification prompt can be seen in Appendix \ref{app:prescreening}, which set definitions, classification criteria and few-shot in-context learning, to classify 4,160 abstracts into the following categories:  \enquote{Supports PTLDS}, \enquote{Supports CLD}, \enquote{Neutral}, \enquote{Unrelated}, or \enquote{Animal Study}. For each classification, the LLM needed to provide an accompanying justification text explaining the reasoning for its classification decision and a confidence level. An example of two abstracts, together with classification and confidence categories, as well as the justification texts, can be seen in Appendix \ref{appendix:WebApplication} in Figures \ref{fig:web_app_demonstration} and \ref{fig:web_app_example}).
The results from this classification step yielded a total of 1,033 abstracts of interest that fell into the target categories of \enquote{Supports PTLDS}, \enquote{Supports CLD} or \enquote{Neutral}.

\subsubsection*{Self-Reflection Classification - Step 2c}

The classifications and the justifications from Step 2b were then reevaluated and reassessed using automated methods and corrected where necessary to improve accuracy. To achieve this, self-reflection prompting technique was used \cite{renze2024self}. Self-reflection prompting is a technique where LLMs re-assess and refine their initial outputs to identify and correct errors, improving overall reasoning and decision-making capabilities. This method has significantly improved problem-solving performance in LLMs \cite{renze2024self}. After executing this step, we found that only 5.3\% (229 out of 4359) of the classifications changed. The largest categories to undergo re-classifications were \enquote{Neutral} (98), where 60\% flipped to \enquote{Supports PTLDS} upon revision, and the \enquote{Unrelated} category (64), where 89\% subsequently became \enquote{Neutral}. Only two abstracts underwent a significant revision from initially \enquote{Supports CLD} to subsequently \enquote{Supports PTLDS}, thus confirming the stability and consistency of the classifications and the approach used. Several examples of abstract classifications and their adjustments in both the final label and justification texts can be seen in Appendix \ref{app:selfreflection}.

\subsection{Human Validation - Step 3}

Following the final classifications, we assessed their reliability and their overall alignment with human expert judgments. To accomplish this, we conducted a comprehensive Inter-Rater Reliability (IRR) analysis~\cite{cohen1960coefficient,fleiss1971measuring}. We sought to compare and establish the degree of agreement between the classifications (1) of our chosen LLM and those of two subject-expert raters, (2) between the two human raters, and (3) between all classifications, including six additional LLMs. Therefore, this evaluation included both pairwise agreement (using Cohen’s Kappa~\cite{cohen1960coefficient}) and multi-rater agreement (using Fleiss’ Kappa~\cite{fleiss1971measuring}). Ultimately, the goal was to determine whether the LLM-based classifications exhibited sufficient reliability comparable to human expert judgement, which would validate the methodology and the experimental results.
The six additional cutting-edge and most advanced LLMs to date included in this validation were Google's Gemini 2.0 Flash (plus the model in the Thinking Mode), Anthropic's Claude 3.5 Sonnet, DeepSeek's R1 model, Alibaba's Qwen 2.5 Max, X's Grok-3. The key metric used to measure inter-rater agreement between two raters while adjusting for chance agreement was Cohen’s Kappa \cite{cohen1960coefficient}, defined as:

\begin{equation}
\kappa = \frac{P_o - P_e}{1 - P_e}
\end{equation}

where \( P_o \) is the observed agreement, and \( P_e \) is the expected agreement by chance. It ranges from -1 (complete disagreement) to 1 (perfect agreement), with standard interpretation thresholds \cite{landis1977measurement} from poor ( \( \kappa < 0.00 \) ) to almost perfect ( \( \kappa \geq 0.80 \) ).
Fleiss’ Kappa  generalises Cohen’s Kappa to multiple raters, providing a single measure of agreement across all LLMs and human raters.
Higher kappa values (max 0.8) indicate strong agreement, reinforcing the validity of automated classification. Low values (< 0.4) suggest significant inconsistencies, highlighting areas of marked divergence.

\subsection*{Agreement Between Human Raters and LLMs}

To establish a baseline for IRR, we first assessed agreement between two domain-expert human raters from the author team. Prior to classification, both raters underwent training to ensure consistency in applying the classification framework (see stance-framing classification prompt in Appendix \ref{app:prescreening} ). This entailed meetings and the provision of a classification training pack. The human raters were provided with the same prompts, criteria, and assumptions used by the LLMs, along with examples of their classifications for reference.
A custom web application was developed for independent annotation (see Appendix \ref{appendix:WebApplication}, Figures \ref{fig:web_app_demonstration} and \ref{fig:web_app_example}), which enabled raters to classify a random sample of 150 abstracts. For each abstract, the raters selected a classification and chose between two possible justification options for their most suitable decision.
Cohen’s Kappa ($\kappa = 0.501$) indicated moderate agreement, aligning with established IRR benchmarks in literature \cite{mchugh2012interrater, viera2005understanding, sim2005kappa} and prior studies in subjective classification tasks such as qualitative content analysis, medical diagnoses, and thematic coding \cite{hallgren2012computing}. Perfect agreement is rarely expected in complex classification tasks due to interpretive differences and variations in emphasis on textual elements. 

Next, we evaluated the agreement between the final revised classifications (produced via GPT-4o-mini’s self-reflection process referred to below as 'GPT') and a set of alternative LLM-generated classifications (\texttt{Gemini}, \texttt{Gemini-Thinking}, \texttt{Claude}, \texttt{DeepSeek}, \texttt{Qwen}, \texttt{Grok-3}), alongside the original LLM classification and those of the two human raters. Table~\ref{tab:cohen_kappa_results} presents Cohen’s Kappa for each pairwise comparison.

\begin{table}[h]
    \centering
    \renewcommand{\arraystretch}{1.2}
    \setlength{\tabcolsep}{10pt}
    \sisetup{round-mode=places, round-precision=3}
    \fontsize{10pt}{12pt}\selectfont
    \begin{tabular}{l S[table-format=1.3]}
        \toprule
        \textbf{Comparison} & \textbf{Cohen's Kappa} \\
        \midrule
        GPT  vs. Original GPT Classification & 0.767 \\
        GPT vs. Gemini & 0.717 \\
        GPT vs. Gemini-Thinking & 0.608 \\
        GPT vs. Qwen & 0.600 \\
        GPT vs. Claude & 0.592 \\
        GPT vs. Human Interrater 1 & 0.583 \\
        GPT vs. Human Interrater 2 & 0.508 \\
        GPT vs. DeepSeek & 0.475 \\
        GPT vs. Grok-3 & 0.458 \\
        \midrule
Human-Human Agreement & 0.501 \\
        \bottomrule
    \end{tabular}
    \caption{Cohen’s Kappa agreement between GPT revised classifications after self-reflection and all other raters/classifications}
    \label{tab:cohen_kappa_results}
\end{table}

Key observations emerge from validation results in Table~\ref{tab:cohen_kappa_results}:

\begin{itemize}
    \item The strongest agreement occurred between the original and GPT's revised classifications ($\kappa = 0.767$), indicating that the LLM self-reflection process refined classifications to a limited degree but did not fundamentally alter them. This suggests that initial classifications were internally consistent, requiring only minor adjustments. This, therefore, supports the claim that there is evidence of an underlying \textbf{stability} in the approach used, confirming both the leveraging of LLMs for this task and the suitability of the designed prompts.
    \item Human vs. GPT's Revised Classification: The agreement between GPT's revised classifications and human raters ($\kappa = 0.583$ for Interrater 1, $\kappa = 0.508$ for Interrater 2) is comparable to human-human agreement ($\kappa = 0.501$). This means that the GPT's revised classifications are at least as consistent with human judgment as human raters are with each other, thus reinforcing the \textbf{validity} of LLM's classifications. Additionally, on the subset of abstracts where both human raters agree, their IRR with GPT's classification was very high at $\kappa = 0.709$. Notably, with respect to LLM model choices, we found that the highest IRR values for both human raters were with GPT and Original GPT classifications ($\kappa = 0.583$ and $\kappa = 0.538$), thus also confirming the \textbf{suitability} of the chosen type of LLM for the classification tasks. 
    \item The highest agreement with an alternative model occurred with \texttt{Gemini} ($\kappa = 0.717$), followed by \texttt{Qwen} ($\kappa = 0.600$). This suggests that these models exhibit classification patterns closest to the GPT's revised outputs, likely due to similarities in training data or reasoning heuristics. 
    \item The lowest agreement is observed with \texttt{Grok-3} ($\kappa = 0.458$) and \texttt{DeepSeek} ($\kappa = 0.475$), indicating notable divergences in their classification outputs. These discrepancies may reflect different conceptual representations of Lyme disease controversies across models and differences in the underlying datasets used for their training, suggesting that these models are not the best candidates for this task.
\end{itemize}

For completeness, to assess global agreement among all human and automated classifiers, we computed Fleiss’ Kappa ($\kappa = 0.537$). This result indicates moderate agreement, aligning with the baseline human-human agreement levels and again reinforcing the consistency of the classification framework as a whole. 
Fleiss’ Kappa further validates the LLM-assisted classification methodology by showing that the aggregate classification structure remains coherent despite inherent variability across all models. This aligns with prior research in natural language processing and thematic coding, where moderate agreement is a reasonable outcome in subjective classification tasks \cite{sim2005kappa, hallgren2012computing}.  
Given the comparable and even higher IRR values between human raters and the GPT's classifications concerning agreement values between the human experts, the IRR analysis confirmed the reliability and methodological rigour of our classification process and the underlying prompts. The alignment between human experts, LLM classifiers, and self-refined classifications demonstrates that LLM-assisted classification of stance or framing in the abstracts is systematic and replicable.

Finally, we also examined the validity of the justification texts generated by the LLM to provide a rationale for the classifications given to each abstract. The analysis found that the human raters achieved a Cohen's Kappa of $\kappa = 0.61$, while both human raters scored $\kappa = 0.71$ against the LLM, again demonstrating that the human raters agreed to a higher degree with the LLM's outputs, then with each other's choices.

\subsection{Thematic Analysis - Step 4}
\label{thematic}

\subsubsection*{Identification and Classification of Overarching Themes}

After all the abstracts were classified into predefined stance categories (\textit{Supports PTLDS}, \textit{Supports CLD}, and \textit{Neutral}), resulting in 1,033 abstracts, a hybrid computational and a theoretically informed thematic analysis overseen by subject experts was conducted to extract deeper conceptual insights into the dominant lines of reasoning and underlying social dynamics within the Lyme disease controversy. This analysis aimed to identify recurrent patterns within the justifications for classification and the abstracts themselves, ensuring a systematic, reproducible, and theoretically grounded approach to the study of contested medical narratives consistent with established social science methodologies for discourse analysis.

Step 4 in Figure \ref{fig:meth} depicts this multi-stage, iterative thematic identification process we employed for combining automated reasoning using multiple LLMs with structured reconciliation and expert human validation that balances computational scalability with interpretive depth. The hybrid approach enabled us to leverage the strengths of both computational pattern recognition at scale and qualitative interpretation of LLMs together with that of the critical human review of subject experts, while integrating refinement and validation in the process to ensure social science validity and conceptual richness. This approach acknowledges that while LLMs can efficiently process large volumes of text, human expertise remains crucial for interpreting social meaning and ensuring theoretical coherence within complex, contested domains like medical controversies.

\subsubsection*{Methodological Approach Description}

\noindent The thematic identification process proceeded in four structured and iterative phases, aiming to ensure methodological rigour, transparency, and conceptual validity, drawing upon established qualitative thematic analysis principles \cite{braun2006using, schreier2012qualitative}:

\begin{enumerate}

    \item \textbf{Multiple-Thematic Identification (Step 4a):}
   \begin{itemize}
        \item  A textual dataset was compiled to automate the theme identification and extraction from the corpus, comprising justification texts from the abstract classification phase in Steps 2b and 2c.  Due to context window limitations inherent in the LLM models, the dataset was reduced to  a random sample of 800 (out of 1,033) justification texts. This sample size was the maximum feasible for effective LLM processing within the technical constraints while providing a substantial and representative corpus for thematic exploration. 
        
        \item  Multiple, diverse and most powerful reasoning LLMs were used for theme identification. This ensured independence in theme identification and mitigated potential biases inherent in any single LLM. Three advanced reasoning LLMs, —\texttt{GPT preview-o1}, \texttt{Gemini 2.0 Flash}, and \texttt{DeepSeek R1}—were each separately tasked with identifying overarching themes within the dataset. 
        
        \item Each LLM was instructed to identify overarching themes without being constrained to a predefined number of themes to extract. This allowed LLMs freedom to enable thematic structures to emerge organically from the data. LLMs were merely instructed to identify \enquote{half a dozen or more overarching themes} based on clustering semantic patterns and recurrent arguments within the text, mimicking a human researcher's inductive thematic coding process.

        \item This process produced three sets of independent themes identified by each LLM, requiring reconciliation.        
    \end{itemize}

    \item \textbf{Thematic Consolidation (Step 4b):}
   \begin{itemize}
        \item  Remarkably, despite operating independently and with different underlying architectures, all three models converged on exactly eight overarching themes (Table \ref{tab:theme_mapping}), suggesting internal consistency in the underlying thematic structures of the Lyme disease discourse and reinforcing the robustness and potential validity of the results beyond any single model's idiosyncrasies.

\begin{table}[htbp]
    \centering
    \renewcommand{\arraystretch}{1.2}
    \setlength{\tabcolsep}{8pt}
    \fontsize{8pt}{10pt}\selectfont
    \begin{tabular}{p{4cm} p{4cm} p{4cm}}
        \toprule
        \textbf{GPT preview-o1 } & \textbf{Gemini 2.0 } & \textbf{DeepSeek R1 } \\
        \midrule
        Persistence vs. Resolution of Infection & Persistence vs. Resolution of Infection & Etiological Mechanisms: Persistent Infection vs. Post-Infectious Immune Responses \\
        \hline
        Diagnostic Uncertainty and Misdiagnosis & Diagnostic Uncertainty and Misdiagnosis & Diagnostic Complexity and Biomarker Development \\
        \hline
        Effectiveness of Antibiotic Therapy & Effectiveness of Antibiotic Therapy & Clinical Management Controversies \\
        \hline
        Role of Immune Dysregulation & Immune Dysregulation & Autoimmune Pathways and Residual Antigenic Debris \\
        \hline
        Psychological vs. Biological Basis & Neurocognitive and Neuropsychiatric Manifestations & Long-Term Outcomes and Symptom Heterogeneity \\
        \hline
        Subjectivity vs. Objectivity of Symptoms & Patient-Centered Experiences & Advocacy and Psychosocial Burden \\
        \hline
        Sociocultural and Ethical Factors & Sociocultural and Ethical Factors & Sociocultural and Institutional Influences \\
        \hline
        Mechanisms of Pathogen Persistence & Mechanisms of Pathogen Persistence & Bacterial Pathogenesis and Host Interactions \\
        \bottomrule
    \end{tabular}
    \caption{Summary of themes independently derived from three LLMs and their mapping.}
    \label{tab:theme_mapping}
\end{table}

        \item  To establish face validity, two co-authors, possessing subject-expertise in Lyme disease controversies and social science discourse analysis, independently reviewed the initial LLM-generated themes. This review phase aimed to ensure the themes resonated with existing knowledge of the Lyme debate and relevant social science concepts. The reviewers investigated conceptual overlaps and divergences across the three models, noting areas of agreement and disagreement in thematic identification. Consideration was given to comparing emergent themes against existing scholarly literature on medical controversies, contested illnesses, and the specific dynamics of the Lyme disease debate, ensuring alignment with established knowledge. Also, an evaluation was made as to whether themes were mutually exclusive, conceptually distinct, and analytically useful for capturing the key dimensions of the Lyme disease discourse from a social science perspective.

        \item Next, the reconciliation process systematically mapped all significant conceptual domains identified by the initial LLM-generated themes. The results can also be seen in Table \ref{tab:theme_mapping}, which shows the alignment of all themes across the  outputs of each LLM.   The cross-model thematic reconciliation was also performed via an iterative inductive-deductive hybrid approach \cite{braun2006using, schreier2012qualitative}, blending automated pattern recognition with expert human judgment. This consisted of two stages, where we first used the advanced \texttt{GPT preview-o1} LLM to generate a consolidated thematic structure across the three models. This LLM-assisted step provided an initial synthesis, highlighting commonalities and potential redundancies for human review. Next, the LLM-reconciled output was manually examined and validated by the same two subject-expert co-authors. This involved  discussions and critical evaluation to ensure conceptual clarity, eliminate redundancies, and, crucially, refine the themes to align with established discourse frameworks in the social sciences, particularly STS \cite{kuhn1962structure}.

\begin{table}[h!]
    \centering
    \renewcommand{\arraystretch}{1.3} 
    \setlength{\tabcolsep}{10pt} 
    \fontsize{8pt}{10pt}\selectfont
\begin{tabularx}{\textwidth}{
    >{\raggedright\arraybackslash}m{2cm}
    >{\raggedright\arraybackslash}m{4.0cm}
    >{\raggedright\arraybackslash}m{8.5cm}
}
        \toprule
        \textbf{Final Consolidated Theme} & \textbf{Description} & \textbf{Social Science Rationale/Relevance} \\
        \midrule
        Active Infection vs. Post-Infectious Immune Activity & Examines whether persistent symptoms are due to an ongoing \textit{Borrelia burgdorferi} infection or a post-infectious immune response, a central debate in Lyme disease. & \textbf{Epistemological Divide \& Paradigm Clash:} Reflects the fundamental epistemological divide in the Lyme controversy, highlighting competing paradigms of disease causation (biological vs. immunological). Connects to STS concepts of scientific controversy, paradigm clashes, and the social construction of scientific facts. \\

        Diagnostic Complexity and Uncertainty & Investigates challenges in Lyme disease diagnosis, including limitations of serological testing, potential misdiagnosis, and the absence of definitive biomarkers. & \textbf{Social Construction of Diagnosis \& Medical Uncertainty:} Illustrates the social construction of diagnostic categories, the inherent uncertainty in medical knowledge, and the limitations of biomedical reductionism in complex illnesses. Relevant to medical sociology's focus on the patients' experience of diagnostic ambiguity, patient navigation of complex medical systems, and the social impact of contested diagnoses. \\

        Therapeutic Controversies and Antibiotic Efficacy & Addresses the contentious issue of prolonged antibiotic therapy, conflicting treatment guidelines, and the clinical efficacy of alternative interventions. & \textbf{Bioethics \& Medical Pluralism:} Highlights the social and ethical dimensions of treatment decisions in contested illnesses, including the balance between evidence-based medicine, patient autonomy, and the influence of advocacy groups on treatment choices. Relates to bioethics, the sociology of medical practice, and the study of medical pluralism and treatment-seeking behaviours. \\

        Immune Dysregulation and Autoimmune Mechanisms & Explores the hypothesis that post-treatment Lyme disease symptoms may stem from immune dysfunction, autoimmunity, or persistent inflammation rather than active infection. & \textbf{Biomedical Framing \& Shifting Paradigms:} Represents a biomedical framing of persistent symptoms within established immunological paradigms, potentially reflecting a shift away from infection-centric models. Connects to STS analysis of how biomedical frameworks shape research agendas and the legitimation of certain types of medical knowledge over others. \\

        Neurocognitive and Neuropsychiatric Manifestations & Focuses on Neurological and Psychiatric Sequelae Linked to Lyme Disease, such as cognitive impairment, neuroinflammation, and psychiatric symptoms. & \textbf{Psychosocial Impact of Contested Illness \& Stigma:} Underscores the psychosocial impact of Lyme disease, including cognitive and mental health challenges, and how these are understood and contested within the controversy. Relevant to medical sociology and medical anthropology's interest in the patient experience of chronic illness, stigma, and the social construction of mental health in contested medical conditions. \\

        Patient-centred Experiences and Advocacy & Highlights patient narratives, diagnostic challenges, disparities in medical recognition, and the broader psychosocial context of Lyme disease. & \textbf{Patient Agency \& Challenging Medical Authority:} Centres on patient perspectives, highlighting patient agency in navigating contested diagnoses, challenging medical authority, and advocating for recognition and alternative treatments. Connects to patient advocacy studies, sociology of patient experience, and the role of online health communities in shaping health discourses. \\

        Sociocultural and Ethical Factors & Examines the role of advocacy groups, public discourse, legal frameworks, and media representations in shaping Lyme disease perceptions and policies. & \textbf{Social Construction of Health Policy \& Media Influence:} Explicitly addresses the broader sociocultural and ethical dimensions of the Lyme controversy, examining the influence of advocacy groups, media framing, and legal frameworks on shaping health policy and public understanding. Directly relevant to SSM's focus on the social determinants of health, health policy analysis, and media studies of health controversies. \\

        Mechanisms of Pathogen Persistence and Biofilm Formation & Investigates microbial survival mechanisms, including persister cells and biofilms, which may contribute to treatment resistance and chronic symptoms. & \textbf{Marginalised Biomedical Research \& Alternative Paradigms:} Represents a more marginalised biomedical research perspective, often associated with CLD advocacy, focusing on mechanisms that challenge mainstream views of pathogen eradication and treatment failure. Connects to STS analysis of scientific marginalisation, the sociology of scientific knowledge production in contested fields, and the dynamics of alternative medical paradigms. \\
        \bottomrule
    \end{tabularx}
    \caption{Key overarching themes extracted from the dataset of abstracts, their classifications, and their definition and grounding in the Science and Technology Studies framework.}
    \label{tab:lyme_themes_rationale}
\end{table}

        \item The final task in Step 4b was to develop a consolidated thematic framework, presented in Table~\ref{tab:lyme_themes_rationale}, which defines each theme alongside its theoretical grounding in social science. This framework aimed to integrate insights from STS, illustrating how social, cultural, and political dynamics shape scientific knowledge production and interpretation, extending beyond biomedical perspectives. By embedding established social science perspectives on Lyme disease discourse, the framework sought to reflect the layered complexity of the controversy, its epistemological tensions, and its broader societal implications, ensuring alignment with existing literature on medical controversies and contested illnesses. To ensure methodological rigour and social science validity, the thematic classification process was explicitly grounded in discourse analysis, the sociology of medical knowledge, and medical controversy research \cite{clarke2007grounded, brown2004embodied, jasanoff2001handbook}. This process systematically integrated LLM-based pattern recognition with expert qualitative interpretation, ensuring that biomedical and sociocultural narratives were meaningfully captured while maintaining conceptual coherence. The hybrid approach preserved the interpretability of computational classifications within broader social and epistemological contexts, ensuring that automated analyses were methodologically sound and aligned with expert human judgment in the study of contested medical knowledge.

    \end{itemize}

    \item \textbf{Abstract Thematic Labelling (Step 4c):} With the themes identified, the last classification task was conducted, to classify all 1,033 abstracts and their classification justifications into the derived themes, which could then aid in conducting a deeper analysis. Automation was once again used for this. \texttt{GPT-o1-mini} was tasked with assigning each abstract/justification pair two most suitable thematic categories. The methodological choice was made recognising the multidimensional nature of Lyme disease discourse and acknowledging the potential for abstracts to address multiple thematic dimensions simultaneously.

    \item \textbf{Expert Validation (Step 4d):}  Finally, we selected a random sample of 50 abstract/justification pairs comprising 100 thematic classifications from Step 4c and asked subject experts to validate the assignment of themes that rounded off the last human-in-the-loop verification of the thematic classification before we proceeded to analysis. In this step, the human validators assessed the classifications for errors. Despite the inherent subjectivity of the task and interpretative overlaps, the human evaluators agreed with 96\% of thematic assignments, thus confirming the validity of this process as a whole.

\end{enumerate}

\section{Results}

\begin{figure}[htbp]
    \centering
    \includegraphics[width=0.75\linewidth]{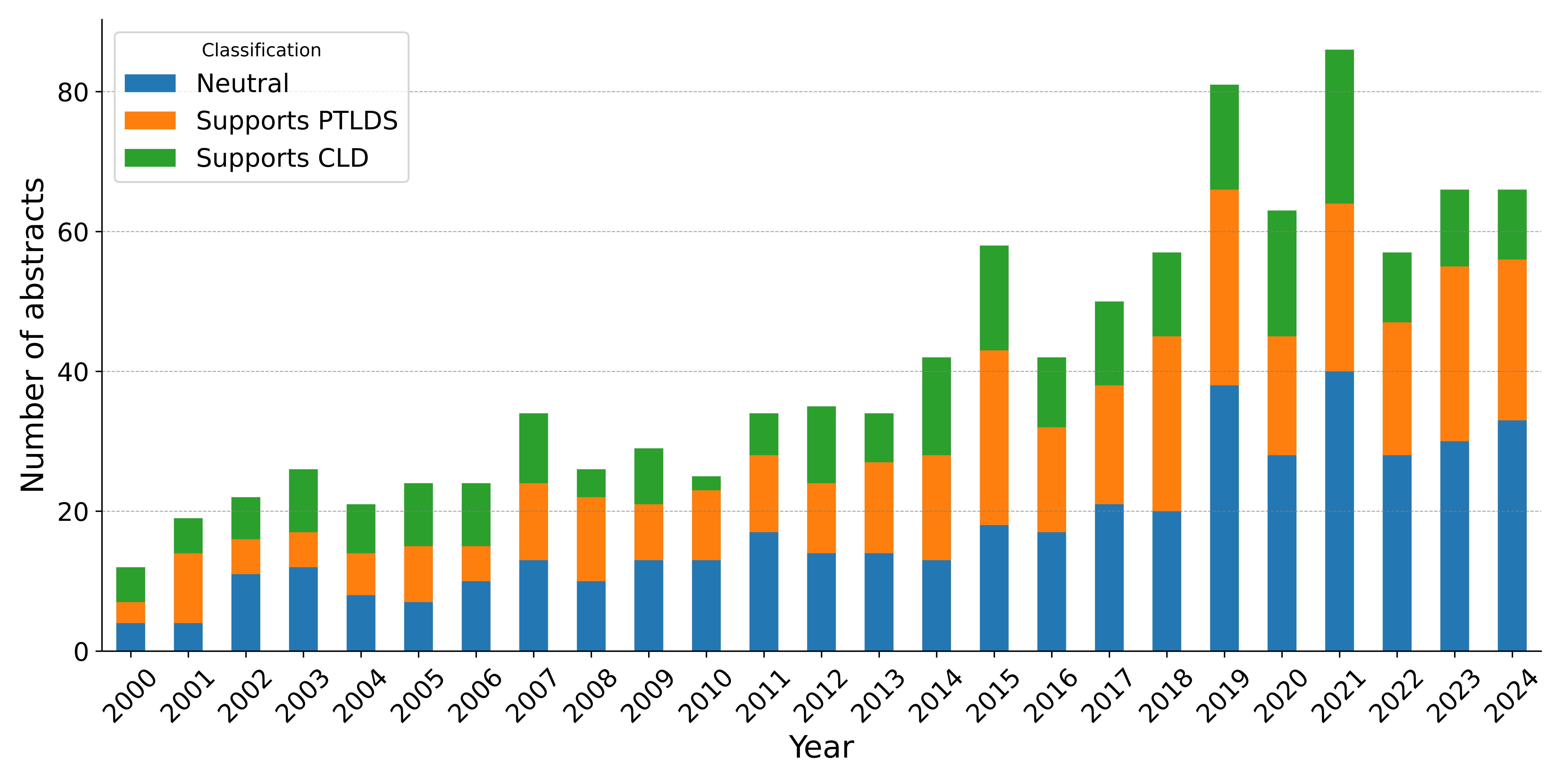}
    \caption{Number of relevant studies by year (2000-2004) and their classification. Classifications consisted of Neutral (blue), Supports PTLDS (orange), and Supports CLD (green).}
    \label{fig:papers_by_year}
\end{figure}

Figure~\ref{fig:papers_by_year} visualizes the yearly distribution of stance classifications, confirming a notable increase in publication volume over time, particularly after 2014, with perceivable surges in 2015, 2019, and 2021. This trend, viewed through an STS lens, underscores the growing societal and academic relevance of the Lyme disease controversy, signalling its transformation into a more debated area within biomedical and public health discourse.
The predominance of the \textit{Neutral} stance, consistently the largest category, suggests general epistemic caution or perhaps strategic neutrality among researchers. Given the persistent diagnostic and therapeutic uncertainties or publication prospects of the studies, this may reflect a field-wide hesitancy to endorse polarised positions. While fewer studies explicitly endorse CLD, their consistent presence indicates a sustained, albeit marginalised, counter-narrative challenging mainstream PTLDS frameworks.  The increase in abstracts supporting the PTLDS perspective, especially post-2010, reveals a dynamic evolution of the scientific dialogue, with a shifting centre of gravity towards the PTLDS framework, even as CLD-aligned discourse persists as a significant dissenting voice. Figure~\ref{fig:papers_by_year} thus highlights the entrenched polarisation and the evolving and contested nature of scientific knowledge production within the Lyme debate.

\begin{figure}[hbtp]
    \centering
    \includegraphics[width=0.75\linewidth]{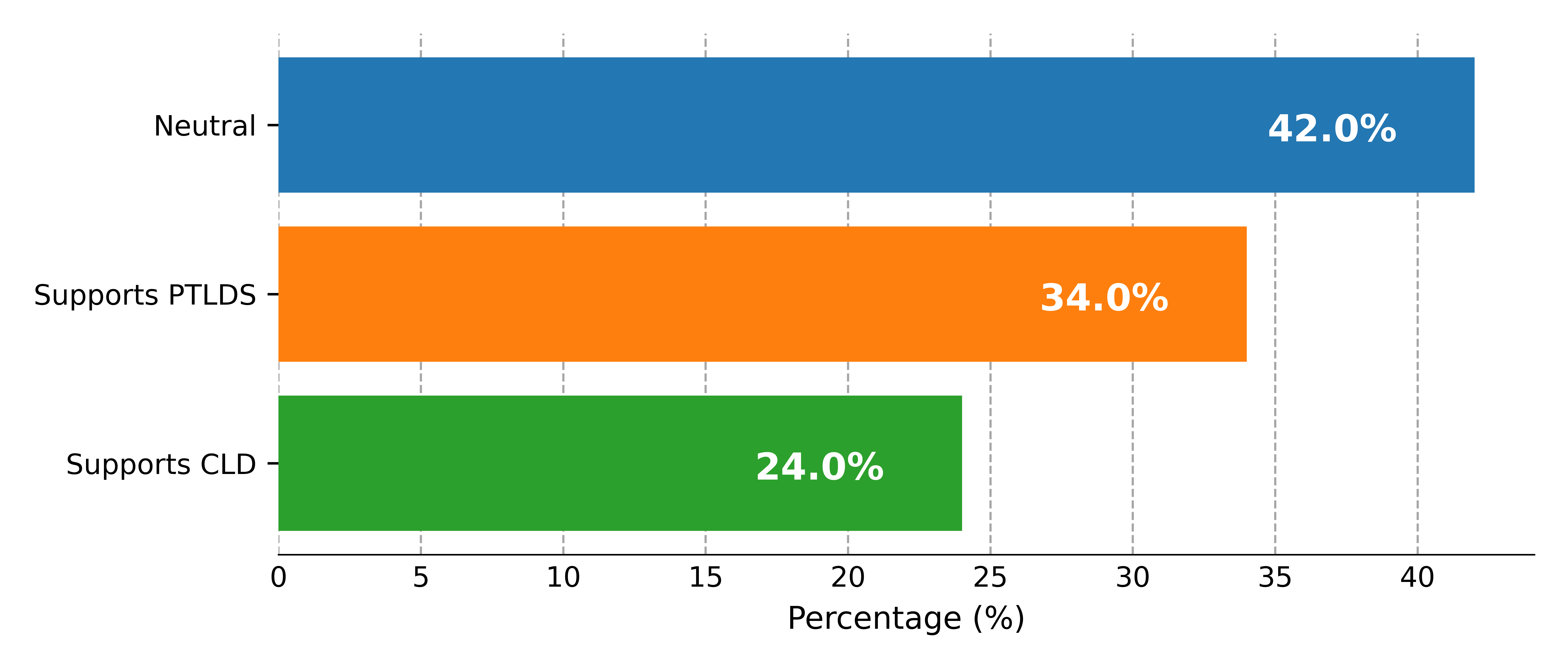}
    \caption{Percentage distribution of classifications. Neutral, Supports PTLDS, and Supports CLD are depicted in blue, orange and green, respectively.}
    \label{fig:percentages}
\end{figure}

Figure~\ref{fig:percentages} confirms these trends by presenting the overall stance distribution across the 25-year dataset. The \textit{Neutral} stance indeed constitutes the largest proportion (42\%), followed by \textit{Supports PTLDS} (34\%) and \textit{Supports CLD} (24\%).  Sociologically interpreted, this aggregate distribution reveals a key structural feature: a substantial portion of the literature strategically navigates the epistemic fault lines of the controversy without fully committing to either pole. However, the combined proportion of PTLDS and CLD-supporting studies (58\%) underscores that the majority of research nonetheless remains deeply polarised, reflecting the enduring scientific and clinical divides within the Lyme disease landscape.

\begin{figure}[hbtp]
    \centering
    \includegraphics[width=0.75\linewidth]{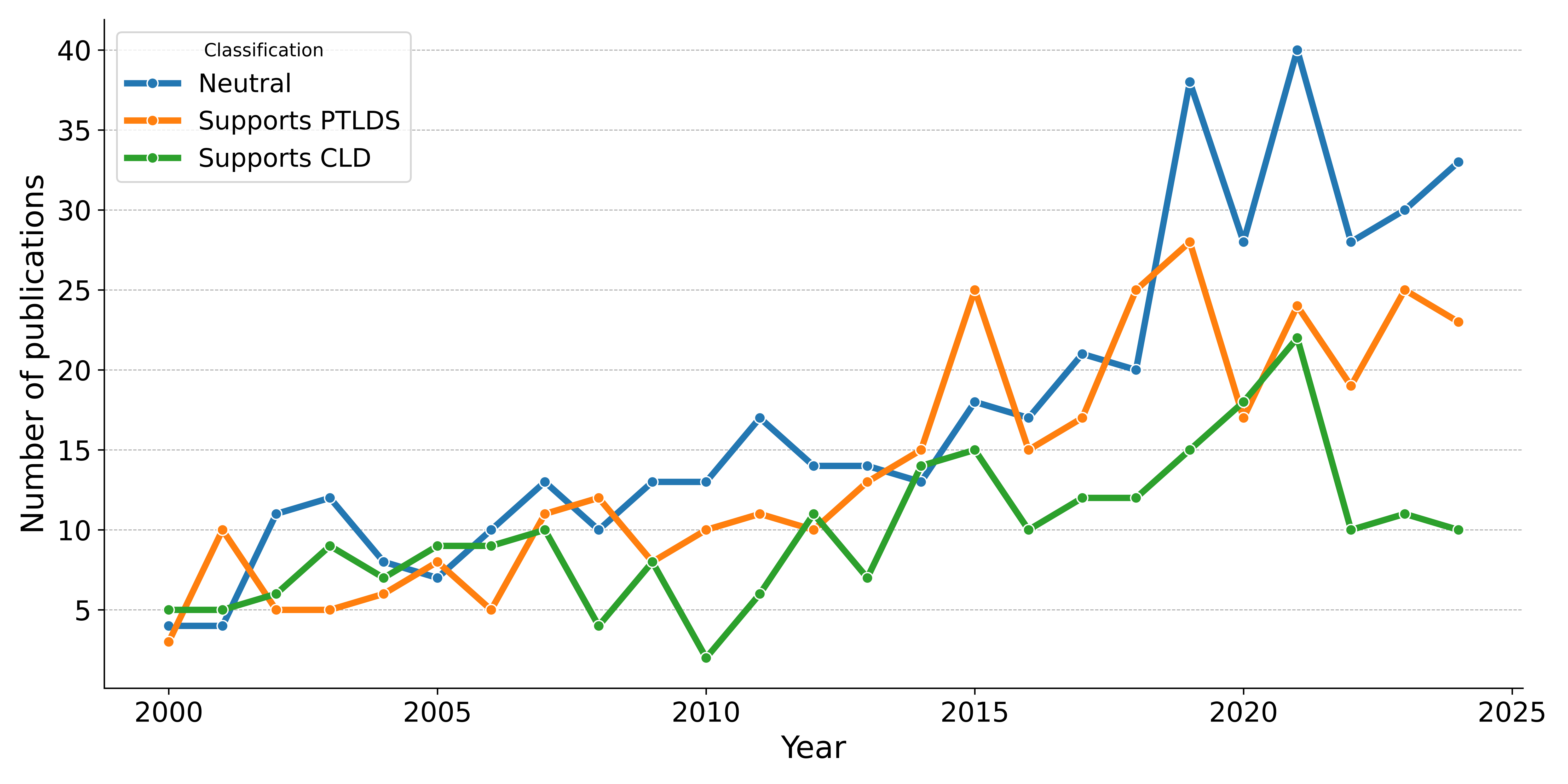}
    \caption{Study classifications on Lyme disease from 2000 to 2025. The yearly count of abstracts on Lyme disease from 2000 to 2025 was classified into three classifications: Neutral, Supports PTLDS, and Supports CLD, which are depicted as blue, orange, and green, respectively.}
    \label{fig:classifications_by_year}
\end{figure}

Figure~\ref{fig:classifications_by_year} provides a more granular and deconstructed view of stance distribution, explicitly showing yearly publication counts per classification and confirming earlier observations. However, Figure~\ref{fig:classifications_by_year} reveals notably that Neutral and Supports PTLDS studies have driven this growth, while Supports CLD publications have remained comparatively flat, even declining recently.  This differential growth, viewed through an STS lens, underscores a shifting centre of gravity in the discourse; whereas the academic conversation expands, increasingly it has centred on Neutral or PTLDS-aligned perspectives. 
The sustained prominence of the Neutral category across years, now visually explicit in Figure~\ref{fig:classifications_by_year}, further highlights strategic epistemic caution within Lyme disease research.  Yearly data show Neutral studies consistently as the largest single category, often exceeding the combined count of polarised stances.  This sustained neutrality, particularly amidst intense debate, likely reflects methodological conservatism where  researchers may be prioritising cautious, evidence-based approaches, avoiding definitive stances given persistent diagnostic and therapeutic uncertainties.

Conversely, the relatively flat trajectory of Supports CLD publications, now clearly differentiated, underscores the persistent marginalization of this perspective. Despite overall research growth, CLD-advocating publications have not seen comparable increases, exhibiting a recent decline.  This visual trend reinforces the interpretation of a field where CLD viewpoints, though consistently present, remain a minority, failing to gain broader traction in mainstream academic discourse. In contrast, Supports PTLDS publications show a pronounced upward trend with a  distinct acceleration, especially post-2014, suggesting the  institutionalisation of the PTLDS framework. This trend indicates a strengthening consensus around immune-mediated explanations and increasing alignment with mainstream medical guidelines.  

\begin{figure}[hbtp]
    \centering
    \includegraphics[width=0.75\linewidth]{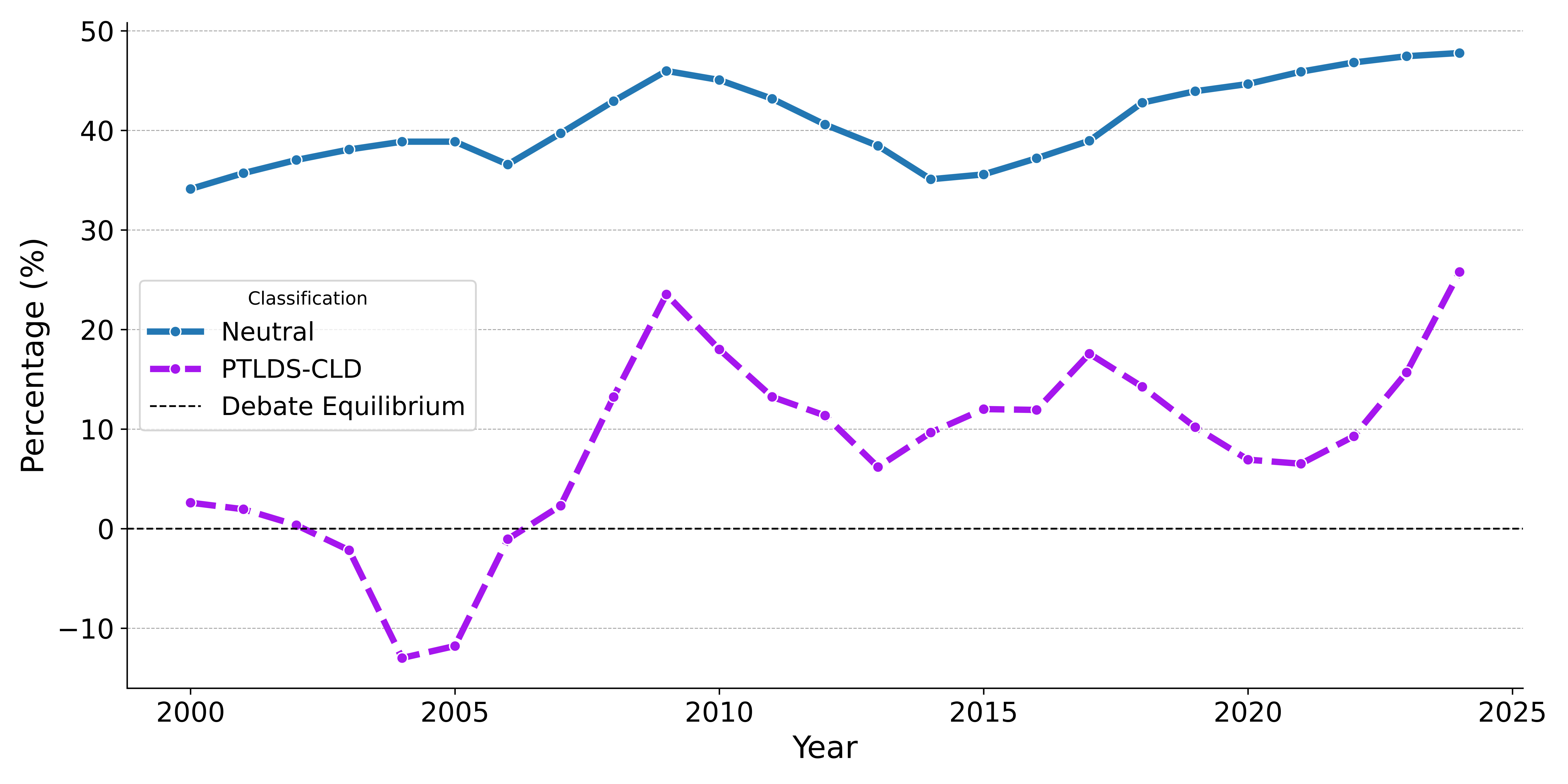}
    \caption{ Smoothed trends in classification percentages by year using the Savitzky-Golay \cite{Savitzky1964smoothing} filter. Lines represent smoothed trends for Neutral and PTLDS-CLD classifications over the period 2000 to 2025.}
    \label{fig:percentage_trends_by_year}
\end{figure}

Figure~\ref{fig:percentage_trends_by_year}, presenting smoothed trends \footnote{We applied the Savitzky-Golay \cite{Savitzky1964smoothing} filter kernel to reduce noise while preserving the underlying signal characteristics, using a second-order polynomial and a window size 10.} in classification percentages, offers a longitudinal perspective on stance evolution.  The solid line confirms the enduring prominence of neutrality as a defining feature of the discourse.  Its upward trajectory, particularly in recent years, reinforces the interpretation of field-wide epistemic caution, reflecting researchers' strategic navigation of persistent uncertainties. 
The dashed line represents the smoothed difference in percentage points between the \textit{Supports PTLDS} and \textit{Supports CLD} classifications. Values above the zero threshold on the figure represent a greater predominance of PTLDS-supporting studies versus CLD and vice versa for negative values. The figure reveals a more dynamic, fluctuating pattern indicative of the shifting balance of power between competing paradigms.  The negative dip in the early to mid-2000s, where CLD-supporting studies held a relative majority, suggests a historically contingent phase of alternative viewpoints gaining temporary traction, potentially fuelled by early patient activism challenging established biomedical narratives.  However, the subsequent and sustained shift into positive territory, accelerating in recent years, once more empirically substantiates the institutionalisation of the PTLDS framework as the increasingly dominant paradigm.  This longitudinal shift signifies a strengthening alignment with mainstream medical consensus and biomedical authority.  While fluctuations persist within the PTLDS-CLD difference, the overall trend underscores the dynamic and continuously evolving nature of the Lyme disease controversy, reflected in shifts in research conclusions and publication volume. Figure~\ref{fig:percentage_trends_by_year} thus provides a macro-level view of these long-term shifts, complementing the granular yearly data presented in Figure~\ref{fig:classifications_by_year} and offering a broader temporal context for understanding the evolving stance landscape.

\begin{figure}[hbtp]
    \centering
    \includegraphics[width=\linewidth]{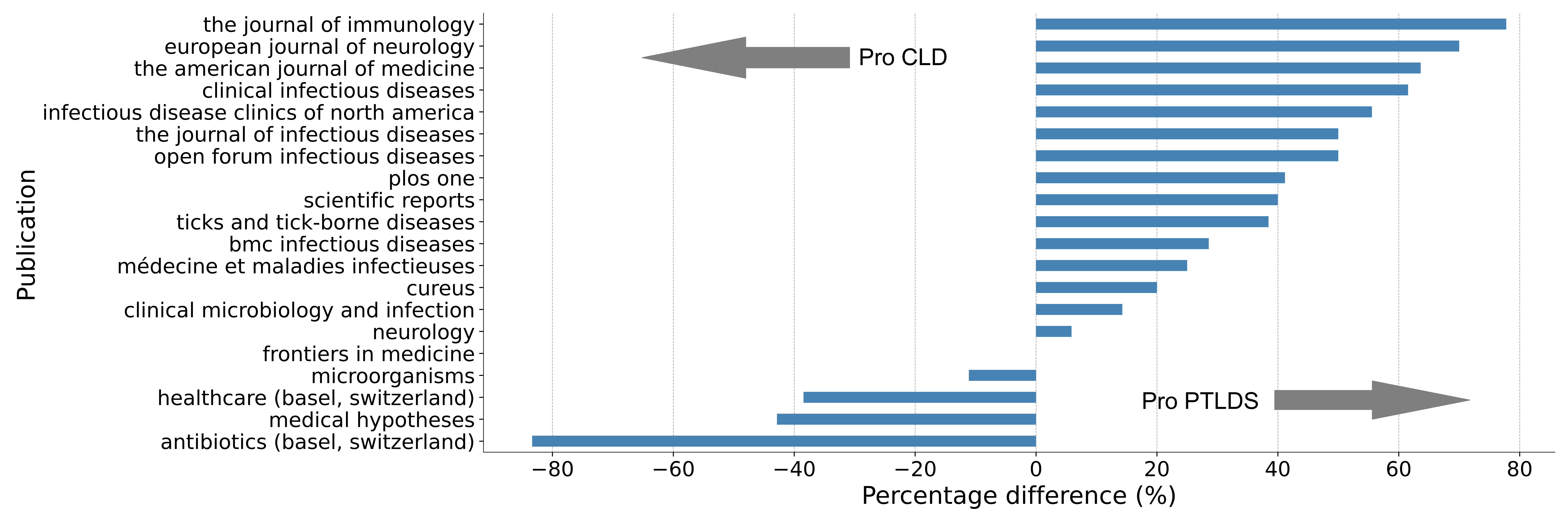}
    \caption{Top 20 journals by volume representing our dataset and depicting their potential bias. Values are the difference in percentage points between published papers classified as supporting PTLDS or CLD.  Positive values indicate a higher proportion of PTLDS-supporting studies for a journal and vice-versa for CLD. }
    \label{fig:journal_preferences}
\end{figure}

Shifting the focus to potentially latent journal-level biases, Figure~\ref{fig:journal_preferences} reveals a stratified epistemic landscape shaped by publication venue. The figure displays the difference in percentage points between \textit{Supports PTLDS} and \textit{Supports CLD} classifications across the top 20 journals by publication volume within our target dataset. Positive values indicate a higher proportion of PTLDS-supporting studies, while negative values suggest a more significant presence of CLD-supporting studies in the given journals' outputs.
This journal-centric view, analysed through STS and Bourdieu's field theory \cite{bourdieu1975specificity} specifically, underscores how a journal's specialisation structures the Lyme controversy, channelling and differentiating the representation of competing knowledge claims.
Leading infectious disease and clinical medicine journals—\textit{Clinical Infectious Diseases}, \textit{The Journal of Infectious Diseases}, and \textit{The American Journal of Medicine}—predominantly publish PTLDS-supporting studies (positive values) with respect to our dataset. This sociologically interpreted proclivity reflects these field-defining venues reinforcing mainstream medical consensus and legitimising the PTLDS framework.  Aligned with established clinical guidelines and institutional authorities like the IDSA, these journals can be seen as functioning as epistemic gatekeepers, preferentially disseminating research congruent with dominant biomedical narratives.

Conversely, journals with broader, hypothesis-driven scopes—\textit{Medical Hypotheses} and \textit{Antibiotics (Basel, Switzerland)}—exhibit a contrasting bias towards CLD-supporting studies (negative values). This suggests that CLD-aligned research, challenging mainstream paradigms, often finds outlets outside core infectious disease venues.  While offering platforms for heterodox perspectives, these journals occupy a more marginalised position within the biomedical field, potentially limiting the broader impact of CLD-supporting research on clinical practice. 
The PTLDS preference in immunology and neurology-focused journals—\textit{The Journal of Immunology} and \textit{The European Journal of Neurology}—further highlights disciplinary influences on potential journal bias, reflecting a research emphasis on immune and neurological dysfunction.  Meanwhile, the slightly more balanced stance in \textit{Ticks and Tick-Borne Diseases} and \textit{BMC Infectious Diseases} likely stems from their broader scope beyond Lyme-specific controversies. Figure~\ref{fig:journal_preferences} thus reveals a segmented epistemic landscape, where journal specialisation and editorial practices actively shape and reinforce distinct perspectives, contributing to the enduring polarisation of the Lyme disease controversy.

To further probe the epistemic influence of publications in Figure~\ref{fig:journal_preferences}, we complement it with the analysis of citation impact data. Our analysis shows that the top 20 most-cited abstracts—a mere 2\% of the dataset—account for 45\% of all citations, thus highlighting a skewed distribution of epistemic power towards a small, highly influential subset of publications.  This concentration, viewed through STS, underscores how a limited number of studies disproportionately shape the discourse's trajectory and impact. 
Analysing stance distribution within these top 20 papers reveals a clear bias:  Supports PTLDS studies dominate (13 of 20), eclipsing Neutral (4) and Supports CLD (3) papers.  Sociologically interpreted, this citation bias suggests a reinforcement of the PTLDS framework as the most impactful paradigm.  The concentration of citations within PTLDS-supporting studies amplifies their visibility, credibility, and perceived scientific authority.
This pattern also extends to the overall citation share across the whole dataset.  While PTLDS-supporting papers constitute 34\% of studies in our dataset, they accrue a disproportionate 52\% of citations.  Conversely, Neutral studies, the largest abstract proportion (42\%), garner a smaller citation share (26\%), and CLD-supporting papers, representing 24\% of abstracts, receive the smallest share (22\%).  This discrepancy between publication volume and citation impact, analysed through the prism of Bourdieu \cite{bourdieu1975specificity}, reveals an epistemic hierarchy.  While Neutral studies represent the largest research output, and CLD-supporting studies maintain a consistent presence, the PTLDS framework commands greater epistemic influence, attracting disproportionate scholarly attention and citations.  Citation analysis thus further substantiates the institutionalisation of the PTLDS framework as the dominant paradigm, shaping not only publication trends but also the perceived impact and influence of Lyme disease research.

\subsection{Thematic Analysis}
To systematically assess the dominant conceptual dimensions in the Lyme disease debate, we categorised each abstract (see Step 4 in Figure \ref{fig:meth}) according to overarching themes (defined in Table \ref{tab:lyme_themes_rationale}) that capture distinct aspects of the controversy. Table~\ref{tab:theme_distribution} presents the thematic distribution of 1,033 classified papers, illustrating their absolute counts and proportional representation. Additionally, it highlights how each theme aligns with the broader debate by indicating the proportion of papers classified as \textit{Neutral}, \textit{Supports PTLDS}, or \textit{Supports CLD}. This analysis provides a structured view of how different scientific perspectives are distributed across key areas of contention.

{\fontsize{8pt}{10pt}\selectfont
\begin{table}[htbp]
    \centering
    \renewcommand{\arraystretch}{1.2} 
    \setlength{\tabcolsep}{10pt} 
    \sisetup{round-mode=places, round-precision=1} 
\fontsize{8pt}{10pt}\selectfont
    \begin{tabular}{r c S[table-format=2.1] S[table-format=2.1] S[table-format=2.1] S[table-format=2.1]}
        \toprule
        \textbf{} & \textbf{} & \textbf{ } & \textbf{Neutral} & \textbf{Supports} & \textbf{Supports} \\        
        \textbf{Theme} & \textbf{Papers} & \textbf{Percent} & \textbf{\%} & \textbf{PTLDS (\%)} & \textbf{CLD (\%)} \\
        \midrule
        Active Infection vs. Post-Infectious Immune Activity & 579 & 56.1 & 11.7 & 49.4 & 38.9 \\
        Diagnostic Complexity and Uncertainty                & 530 & 51.3 & 77.0 & 16.2 & 6.8  \\
        Therapeutic Controversies and Antibiotic Efficacy   & 365 & 35.3 & 20.8 & 40.5 & 38.6 \\
        Neurocognitive and Neuropsychiatric Manifestations  & 196 & 19.0 & 61.7 & 21.4 & 16.8 \\
        Immune Dysregulation and Autoimmune Mechanisms      & 192 & 18.6 & 28.1 & 62.5 & 9.4  \\
        Patient-Centered Experiences and Advocacy          & 149 & 14.4 & 82.6 & 8.1  & 9.4  \\
        Mechanisms of Pathogen Persistence and Biofilm Formation & 30  & 2.9  & 10.0 & 0.0  & 90.0 \\
        Sociocultural and Ethical Factors                  & 25  & 2.4  & 76.0 & 24.0 & 0.0  \\
        \bottomrule
    \end{tabular}

    \caption{Distribution of studies by theme and classifications supporting PTLDS/CLD or neutrality.}
    \label{tab:theme_distribution}
\end{table}
}

Among the themes in Table~\ref{tab:theme_distribution}, \textit{Active Infection vs. Post-Infectious Immune Activity} emerged as the dominant topic  associated with more than half of the studies. This theme reflects the fundamental divide in Lyme disease research, where one faction attributes persistent symptoms to immune dysfunction (Supports PTLDS, 49.4\%). At the same time, the other endorses the hypothesis of bacterial persistence (Supports CLD, 38.9\%). The relatively low proportion of neutral papers (11.7\%) compared to the overall percentage (42\% refer to Figure \ref{fig:percentages}) underscores the polarising nature of this theme, as most studies align explicitly with one of the two competing explanatory models. 

Closely related to this core controversy is the  \textit{Diagnostic Complexity and Uncertainty} theme, which also appears in half of the target studies. In contrast to the previous theme, here there is an overwhelming neutrality of studies in this category (77\%), highlighting the ongoing challenges in establishing clear diagnostic criteria over the last quarter of a century and is a problem that has been exacerbated by the limitations of serological tests and the absence of universally accepted biomarkers \cite{brunton2017stakeholder,aguero2015lyme}. Despite the overwhelming neutrality that has sustained the controversy amidst diagnostic ambiguity, a larger proportion of studies lean towards PTLDS (16\%), while only 6.8\% align with CLD, indicating that, while uncertainty dominates, the prevailing inclination, although modest, favours an immune-mediated rather than infection-driven explanation for persistent symptoms.

Discussions surrounding treatment strategies are equally contentious, as reflected in the  \textit{Therapeutic Controversies and Antibiotic Efficacy} theme, which accounts for around a third of the studies. The classification breakdown reveals a near-even split between PTLDS-supporting (40.5\%) and CLD-supporting (38.6\%) papers, with a significantly smaller than average distribution to a neutral stance, illustrating the ongoing debate over whether extended antibiotic regimens provide a therapeutic benefit or pose unnecessary risks. This division again mirrors the long-standing disagreements between major health organisations regarding treatment guidelines, further reinforcing the unresolved nature of this issue.

Beyond these core debates, several secondary themes provide insights into the broader implications of Lyme disease research and discourse. The theme of \textit{Immune Dysregulation and Autoimmune Mechanisms}, which is identified in a fifth of studies, is predominantly associated with PTLDS (62.5\%), suggesting a growing recognition of immune dysfunction as a plausible driver of persistent symptoms. Conversely, bacterial persistence receives significantly less support within this thematic category (9.4\%), confirming that research into immune dysregulation tends to align more closely with the PTLDS framework than the CLD perspective.

A related but distinct theme, \textit{Neurocognitive and Neuropsychiatric Manifestations}, also emerges in a fifth of the studies. A significant majority of these, however, remain neutral (61.7\%), reflecting the ambiguity surrounding whether neurocognitive symptoms arise from residual infection, immune-mediated mechanisms, or other secondary effects. While PTLDS-supporting papers (21.4\%) slightly outnumber CLD-supporting papers (16.8\%), the relatively balanced distribution suggests that both explanatory models remain viable considerations in this domain in the presence of predominant uncertainty. 
Outside of the biomedical discourse, the role of patient experiences and advocacy remains a notable but underexplored dimension. The theme of \textit{Patient-Centered Experiences and Advocacy} represents approximately an eighth of all studies characterised by overwhelming neutrality (82.6\%). This high neutrality suggests that, while patient narratives are acknowledged, few studies explicitly frame them within either the PTLDS or CLD paradigm. Nonetheless, small proportions of studies equally support PTLDS (8.1\%) or CLD (9.4\%), reflecting the residual tensions between scientific discourse and patient advocacy efforts.

Although the theme of \textit{Mechanisms of Pathogen Persistence and Biofilm Formation} is among the least represented (2.9\% of studies), it stands out for its strong association with CLD (90.0\%). This overwhelming alignment suggests that research on bacterial persistence may primarily be conducted within the CLD framework, reinforcing its position as the central scientific rationale for the chronic infection hypothesis. However, the limited number of papers in this category indicates that, despite its prominence in CLD discourse, empirical investigations into biofilms and persister cells remain relatively scarce. 
Lastly, \textit{Sociocultural and Ethical Factors} constitute the least explored theme, appearing in only 2.4\% of the studies. The high proportion of neutral papers (76.0\%) suggests that while sociopolitical influences are acknowledged, they are seldom the primary focus of scientific inquiry in this controversy. However, within this small subset of studies, PTLDS receives more explicit support (24.0\%) than CLD (0\%), potentially reflecting how institutional and ethical discussions more frequently align with the mainstream medical consensus rather than the alternative chronic Lyme paradigm.

Table~\ref{tab:theme_trends} recasts the above analysis into a temporal frame. The table, therefore, reveals temporal shifts in thematic focus across the past three decades. The thematic analysis from this table identifies shifts in the scientific focus of Lyme disease research and discourse over time, normalising the percentages to account for the incompleteness of the current decade. The most studied theme, \textit{Active Infection vs. Post-Infectious Immune Activity}, has declined from 33\% of papers in the 2000s to 24\% in the 2020s, reflecting a broader transition from infection-centric to immune-mediated explanations. Meanwhile, \textit{Diagnostic Complexity and Uncertainty} have grown (21\% $\rightarrow$ 25\% $\rightarrow$ 29\%), pointing toward continued and unresolved challenges in establishing reliable biomarkers and diagnostic criteria. Similarly, \textit{Therapeutic Controversies and Antibiotic Efficacy}, though declining in prevalence (20\% $\rightarrow$ 20\% $\rightarrow$ 14\%), remains one of the most contested areas as discussed previously.

\begin{table}[htbp]
    \centering
    \renewcommand{\arraystretch}{1.2} 
    \setlength{\tabcolsep}{10pt} 
    \fontsize{8pt}{10pt}\selectfont
    \begin{tabular}{r c c c}
        \toprule
        \textbf{Theme} & \textbf{2000s (\%)} & \textbf{2010s (\%)} & \textbf{2020s (\%)} \\
        \midrule
        Active Infection vs. Post-Infectious Immune Activity & 33 & 28 & 24 \\
        Diagnostic Complexity and Uncertainty                & 21 & 25 & 29 \\
        Therapeutic Controversies and Antibiotic Efficacy   & 20 & 20 & 14 \\
        Neurocognitive and Neuropsychiatric Manifestations  & 11 & 8  & 11 \\
        Immune Dysregulation and Autoimmune Mechanisms      & 10 & 8  & 10 \\
        Patient-Centered Experiences and Advocacy          & 3  & 8  & 10 \\
        Sociocultural and Ethical Factors                  & 1  & 1  & 2  \\
        Mechanisms of Pathogen Persistence and Biofilm Formation & 1  & 2  & 1  \\
        \bottomrule
    \end{tabular}
    \caption{Shifts in the percentage of studies by thematic focus across decades.}
    \label{tab:theme_trends}
\end{table}

These findings highlight distinct thematic patterns in Lyme disease research, reinforcing the entrenched divisions that define the controversy. The central debate over persistent infection versus immune-mediated pathology remains a dominant axis of discourse. Yet, its proportional representation has declined as research has increasingly shifted toward diagnostic complexity and uncertainty. Studies focusing on diagnostic complexity have steadily increased over the decades, now representing the most expanding area of research, underscoring the persistent difficulty in establishing definitive diagnostic criteria. Treatment strategies remain a significant point of contention, though research on antibiotic efficacy has proportionally declined, suggesting a decreased emphasis on therapeutic debates over time. While immune dysregulation and neurocognitive symptoms continue to be actively explored within the PTLDS framework, research on bacterial persistence and biofilm formation remains confined to a niche subset of CLD-focused studies, reflecting its limited expansion within mainstream biomedical discourse. The high neutrality observed in patient-centred and sociocultural discussions underscores the marginal integration of experiential perspectives into scientific debates despite a modest increase in advocacy-related research in recent years. These thematic distributions illustrate the scientific dimensions of the Lyme disease debate and its broader structural divisions where competing frameworks persist across clinical, immunological, and sociopolitical contexts, shaping the discourse, research priorities and institutional positioning.

\section{Discussion}

Prior research has offered crucial insights into the contentious character of Lyme disease by examining patient advocacy, the influence of alternative medical viewpoints, and the effect of scientific ambiguity on public confidence. Nevertheless, these investigations have predominantly been qualitative, depending on discourse analysis, patient trajectory studies, and theoretical frameworks. This research used a hybrid and novel AI-driven technology combined with the human-in-the-middle oversight to identify attitudes and analyse themes in a large dataset spanning 25 years. This methodology enabled a structured examination of the discussion on Lyme disease over time and at scale, offering a quantifiable view of persistent debates that have exceeded prior research endeavours. 

\subsection{Evolution of the Academic Discourse on CLD and PTLDS}

The evolution of Lyme disease research revealed a significant shift in stance distributions over time (\textbf{RQ1}). In the early 2000s, CLD-supporting studies were relatively more frequent, reflecting ongoing debates about the persistence of \textit{Borrelia burgdorferi} post-antibiotic treatment. Yet, after 2010, PTLDS-supporting studies increasingly outnumbered CLD-supporting ones, a trend that has accelerated in recent years. This shift coincided with the consolidation of the PTLDS framework within major medical organisations and an increasing emphasis on immune-mediated rather than infection-driven explanations for persistent symptoms.
Alongside this shift, thematic trends revealed a transition in research priorities. This trend underscored the prolonged difficulty in establishing reliable biomarkers and diagnostic criteria, which remains a key barrier to consensus in the field. Similarly, research on therapeutic controversies, including antibiotic efficacy, has proportionally declined (20\% → 14\%), suggesting a decreasing emphasis on long-term antibiotic treatment trials. This shift may also reflect clinical trial results discouraging prolonged antibiotic use and institutional funding trends prioritising alternative research directions. These findings indicate that while stance polarisation persists, the scientific focus has reoriented from direct treatment interventions toward diagnostic and mechanistic questions. This pattern aligns with more significant changes in discussions about Lyme disease.

This study's primary contribution in quantifying discourse shifts throughout time on this controversy using a large-scale dataset exceeds prior research efforts. Although, investigations by \citet{pascal2020emergence,puppo2023social, hinds2019heterodox} examined the polarisation and impact of conflicting medical viewpoints on Lyme disease discourse, frequently highlighting the significance of social movements, patient advocacy, and media influence, these studies have not incorporated a systematic chronological analysis. Our confirmation that since 2010, there has been a significant rise in research endorsing PTLDS perspectives, especially in high-impact journals accompanied by disproportionate citation volumes and impact, whilst studies supporting CLD have diminished, has provided  empirical evidence of a shift in scientific consensus, which was formerly assumed from limited qualitative evaluations.

\subsection{Influence of Journal Specialisation and Editorial Focus}

Our analysis also supports the notion that journal specialisations shape the discourse on Lyme disease (\textbf{RQ2}), 
where we find that leading infectious disease and clinical medicine journals, predominantly publish PTLDS-supporting studies. This reflects alignment with mainstream clinical guidelines, particularly those endorsed by the IDSA, which maintains that prolonged antibiotic therapy for CLD lacks sufficient evidence.
\citet{uzzell2012whose} and \citet{bloor2021knowledge} illustrated that journalistic framing frequently favours established medical institutions, reinforcing mainstream opinions and marginalising alternative viewpoints.
In contrast, CLD-supporting studies are more frequently published in journals with broader or hypothesis-driven scopes, underscoring the claim that research favouring bacterial persistence and extended antibiotic treatment often finds publication outlets outside high-impact infectious disease journals, potentially limiting its influence on clinical practice. The strong preference for PTLDS-supporting studies in neurology and immunology-focused journals further indicates that PTLDS is increasingly framed as an immunological or neurological disorder rather than an active bacterial infection.
These journal-level biases may contribute to a self-reinforcing effect, wherein PTLDS-supporting research is more likely to appear in high-impact venues, increasing its credibility and citation impact. At the same time, CLD-supporting studies remain within a more restricted epistemic niche. Future investigations could explore whether this distribution reflects methodological quality differences or broader editorial and peer-review preferences that exclude CLD perspectives from mainstream discourse.

Despite the increasing acknowledgement of patient-centred perspectives in Lyme disease discourse, their representation in high-impact journals remains limited. Studies by \citet{baarsma2022knowing, rebman2017clinical} and \citet{singer2019empowerment} highlight the medical scepticism and social stigma experienced by patients diagnosed with persistent Lyme disease symptoms, leading many to seek alternative healthcare solutions. While prior research has documented these qualitative aspects, our study quantifies the growing scholarly emphasis on patient experiences, particularly since the late 2010s. However, this research remains descriptive mainly, with limited integration into clinical or mechanistic frameworks.

\subsection{Thematic Structures and Their Scientific and Institutional Implications}

Our thematic analysis provided novel and deeper insights into framing Lyme disease debates (\textbf{RQ3}) across time. While the infection persistence vs. immune-mediated pathology controversy has remained central, its proportional representation has declined (33\% $\rightarrow$ 24\%) over the decades, suggesting a shift in research emphasis. In contrast, studies on diagnostic complexity and uncertainty have increased (21\% $\rightarrow$ 29\%), highlighting ongoing methodological challenges in standardising Lyme disease diagnostics. 
Therapeutic debates remain polarised but have decreased in prominence (20\% $\rightarrow$ 14\%), reflecting a redirection in focus toward immune and neurological complications rather than antimicrobial persistence. Research on bacterial persistence and biofilm formation remains a niche field, with over 90\% of studies in this category aligning with CLD perspectives. However, this area has not expanded significantly within mainstream Lyme disease research.

Beyond biomedical discourse, research on patient-centred experiences and advocacy has grown (3\% $\rightarrow$ 10\%), reflecting increasing acknowledgment of patient narratives. However, high neutrality within this category (82.6\%) suggests that patient experiences are often studied descriptively rather than integrated into clinical or scientific frameworks. Similarly, sociocultural and ethical discussions have remained a small but expanding domain, underscoring the broader societal implications of the Lyme disease controversy.
Our findings align with prior studies by \citet{pascal2020emergence} and \citet{dumes2020divided}, which examined the role of media in shaping public and scientific perceptions of Lyme disease. While these studies analysed these trends qualitatively, our research quantifies the increasing prominence of patient-centred research relative to infection-focused studies, providing empirical validation for the reciprocal influence between scientific discourse and public narratives.

\subsection{Implications of the Hybrid AI-Computational and Human-in-the-loop Text Analysis Methodology}

This study's hybrid computational-interpretive methodology offers a scalable yet theoretically rigorous approach to systematically analysing large-scale textual datasets in social science research and beyond. This study demonstrated its value for studying complex and contested medical conditions. Combined with structured human oversight and validation grounded in established social science frameworks, it has demonstrated how AI-assisted classification and content analysis can be leveraged without displacing critical human interpretation or compromising reliability. The proposed methodological approach ensured the robustness and validity of replicable results, with the approach transferable to similar domains.

\subsection{Broader Implications and Future Directions}

Beyond specific stance distributions, this research highlights structural and epistemic challenges in Lyme disease. Key considerations include:

\begin{itemize}
    \item The continued diagnostic uncertainty signals a need for better biomarkers and molecular tools to bridge discordant perspectives. 
    \item Reduced emphasis on infection persistence and therapeutic controversies suggests shifting funding and research priorities toward immune-mediated explanations and diagnostic improvements.
    \item The incremental attention to patient advocacy underscores the importance of inclusive research integrating qualitative and patient-reported data.
    \item The limited expansion of bacterial persistence research indicates remaining contention over empirical evidence, funding, or shifting clinical frameworks.
    \item Institutional factors and clinical guidelines shape research agendas. Additional examination of policy, medical guidelines, and funding mechanisms will elucidate how shifts in public health priorities affect the landscape of Lyme disease research.
\end{itemize}

Our study aligns with agenda-setting \cite{mccombs1972agenda}, diffusion of innovations \cite{rogers2003diffusion}, and framing theory \cite{entman1993framing} to illustrate how institutional gatekeepers, clinical norms, and framing tactics shape the discourse, and how scientific disagreements are influenced by communication and institutional factors, offering a new viewpoint on the underlying reasons for ongoing Lyme disease challenges by applying these theories. This research also underscores the interplay between scientific authority, patient autonomy, and public health communication for disputed medical conditions \cite{ciotti2023three,bloor2021knowledge}.

Overall, our large-scale, AI-assisted analysis deepens and extends prior qualitative findings, emphasising the sociopolitical factors that render Lyme disease a complex, evolving controversy. These insights reinforce the need for policy interventions supporting equitable research funding, patient-centred care, and improved diagnostic clarity. Future research should bridge biomedical consensus with authentic patient experiences, employing AI-driven methodologies to detect emerging patterns and improve health communication strategies. The persistence of unresolved debates around Lyme disease calls for a balanced approach integrating scientific rigour with empathetic, patient-focused practice.

\subsection{Limitations}

Several limitations must be acknowledged. First, the paper classification relied on abstracts, which may not fully capture nuanced positions or methodological details that may have been more clearly expressed in the full texts of the studies. The dataset was limited to indexed, peer-reviewed literature, excluding unpublished and non-English studies, potentially biasing stance distributions. 
The IRR findings highlighted some degree of  variability in reasoning and outputs between the LLMs and human experts. It is possible that ensembling models' outputs and aggregating their classifications instead of relying on only one LLM could have enhanced classification reliability further.
Journal-level biases may suggest editorial influence on discourse, but this study does not account for factors such as funding sources or peer review dynamics that may shape publication trends. Thematic categorisation, while informative, may oversimplify overlapping topics and emerging interdisciplinary perspectives - some abstracts may have spanned more than two themes. Additionally, stance shifts over 25 years may reflect evolving clinical definitions rather than changes in underlying scientific evidence.
This study has mapped the structure of Lyme disease research but does not assess the empirical validity of competing claims.

\section{Conclusion}
Over the past quarter-century, the academic discourse surrounding Chronic Lyme Disease (CLD) and Post-Treatment Lyme Disease Syndrome (PTLDS) has become a deeply entrenched and complex controversy, fuelled by scientific disputes, patient activism, and pervasive media narratives.  This study offers an unprecedented, large-scale quantification of this epistemic landscape  through a rigorous and innovative hybrid AI-driven methodology. Our research transcended the limitations of traditional qualitative approaches for content analysis by analysing thousands of scholarly studies spanning this period, providing a temporally sensitive and granular understanding of the evolving dynamics of this enduring medical debate.  The core novelty lies in our synergistic combination of cutting-edge large language models with structured human validation, enabling a level of systematic and scalable discourse analysis previously unattainable.

With our novel hybrid approach, our analysis revealed a significant and consequential shift in the academic conversation on this controversy, indicating that a clear transition from infection-centric models of Lyme disease towards immune-mediated explanations for persistent symptoms has occurred.  This trend is particularly pronounced within high-impact clinical and immunology journals, signalling a growing institutional endorsement of the PTLDS perspective.  Conversely, research supporting CLD, while consistently present, has remained largely relegated to hypothesis-driven publications, highlighting a persistent epistemic asymmetry within mainstream biomedical discourse.  This divergence powerfully underscores the impact of journal specialisation and potential editorial biases in shaping competing scientific perspectives' dissemination and perceived validity.  Furthermore, our AI-driven thematic analysis showed the enduring centrality of diagnostic complexity and therapeutic controversies, even as the dominant explanatory models and research priorities undergo significant transformation, demonstrating the robustness of our methodology in capturing nuanced thematic evolutions.

The key contribution of this study has been twofold: first, the provision of empirically grounded, large-scale insights into the structural and epistemic dynamics of the Lyme disease controversy, and second, the establishment of a robust and replicable hybrid AI-driven methodology for analysing complex medical debates.  This innovative approach, demonstrably effective in dissecting the Lyme disease discourse, offers a powerful new tool for social science research in other fields.  Our findings have direct implications for policymakers, clinicians, and communication strategists.  For policymakers, they highlight the urgent need for equitable research funding policies that actively promote diverse scientific inquiry and challenge epistemic marginalisation.  The persistent diagnostic ambiguity, revealed with unprecedented clarity through our AI-driven analysis, necessitates a renewed commitment to establishing standardised diagnostic criteria.  For clinicians, our study reinforces the critical importance of adopting patient-centred care models that validate lived experiences, irrespective of diagnostic uncertainties.  For communication strategists, we underscore nuanced and empathetic communication strategies are pivotal in bridging the persistent divide between scientific consensus and patient narratives.

\bibliographystyle{unsrtnat}

\appendix

\clearpage

\section{Classification Prompts and Example Outputs}
\label{app:prescreening}

Below is the prompt used for pre-screening classification GPT-4o-mini to execute the Step 2a of the methodology depicted in Figure \ref{fig:meth}.

\begin{mdframed}[backgroundcolor=gray!10]
\fontsize{8pt}{10pt}\selectfont
\begin{verbatim}
You are a world-leading expert in medical literature, specialising in Lyme disease and the controversies
surrounding chronic Lyme disease (CLD) and post-treatment Lyme disease syndrome (PTLDS).

Your task is to evaluate a single scientific paper abstract at a time to determine if it has 
relevance to either CLD or PTLDS for further processing.

Definitions:
- Chronic Lyme Disease (CLD): A contested condition wherein some believe symptoms result from an ongoing 
Borrelia burgdorferi infection is often cited as requiring extended antibiotic therapy.
- Post-Treatment Lyme Disease Syndrome (PTLDS): Refers to persistent symptoms (e.g., fatigue, pain, 
cognitive issues) following standard Lyme disease treatment, generally understood not to involve active
infection.

Categories:
Each abstract should be categorised into one of the following three categories:

1. Potentially Related to CLD/PTLDS:
   - Select this category if the paper involves human studies and contains explicit or implicit references, 
   findings   or terminology potentially related to CLD or PTLDS. This includes abstracts where the 
   connection to CLD or PTLDS is indirect, uncertain, or requires additional processing to 
   determine relevance.

2. Definitely Unrelated:
   - Select this category if the study has no relevance to CLD or PTLDS in human subjects. 
   Abstracts discussing other medical conditions and unrelated tick-borne diseases.

3. Animal Study:
   - Select this category if the study is entirely focused on animal subjects with no direct relevance to 
   human cases of CLD or PTLDS.

Input:
For each abstract, you will receive an index number, a paper title, and the abstract text.

JSON Output Structure:
Your output must be a JSON object containing only the evaluation of the current abstract. Each response 
should include the following fields:

- `index`: The index number of the abstract, which MUST match the input.
- `classification`: One of "Potentially Related to CLD/PTLDS," "Definitely Unrelated," or "Animal Study."
- `confidence`: Confidence level in the classification ("High," "Medium," or "Low").

Example Output:
{
  "index": 700,
  "classification": "Potentially Related to CLD/PTLDS",
  "confidence": "High"
}

ABSTRACT:
< input >
\end{verbatim}
\end{mdframed}

Following the above classification, stance-framing detection was then executed as outlined in Step 2b of the methodology in Figure \ref{fig:meth} with the abridged prompt below, where the three abstracts provided to the LLM as part of few-shot learning are omitted for brevity, but can be found on the Github repository of the project:

\begin{mdframed}[backgroundcolor=gray!10]
\fontsize{8pt}{10pt}\selectfont
\begin{verbatim}

You are a world-leading expert in medical literature analysis and particularly in medical literature, 
specialising in Lyme disease and the debates surrounding chronic Lyme disease (CLD) and post-treatment 
Lyme disease syndrome (PTLDS). Your task is to evaluate a series of scientific paper abstracts related 
to Lyme disease and determine each abstract's stance on PTLDS and CLD. This task requires scientific 
impartiality to minimize any bias in classifying abstracts concerning disputed subjects such as CLD or 
PTLDS. The current literature on this topic is polarised and can be expressed as follows: 

"Some medical experts argue that post-treatment symptoms, experienced by a subset of patients after 
completing standard antibiotic therapy, can be attributed to what they define as Post-Treatment Lyme 
Disease Syndrome (PTLDS). According to some health organisations, these symptoms—ranging from fatigue 
to cognitive impairment—are likely caused by immune responses or tissue damage, rather than persistent 
infection. PTLDS concerns patients who experience persistent symptoms for at least six months after the 
completion of recommended treatment for Lyme disease.

Conversely  , other organisations advocate for the recognition of chronic Lyme disease (CLD), contending 
that ongoing infection or immune dysfunction may be responsible for these symptoms and recommend extended 
antibiotic regimens, pointing to contested evidence of patient improvement. CLD is a broader term that 
encompasses a range of unexplained symptoms that patients attribute to a persistent Lyme disease infection 
that survives recommended treatment.  Both PTLDS and CLD involve persistent symptoms, such as fatigue, 
pain, and neurological issues, and both face with diagnostic difficulties due to nonspecific symptoms 
and lack of clear biomarkers, contributing to ongoing debates in the medical community."

Based on the above context and the debate, your task is to classify each abstract and determine each 
abstract's explicit or implicit stance regarding CLD or PTLDS based on the following classifications:

- Supports PTLDS
- Supports CLD
- Neutral
- Unrelated
- Animal Study

Definitions for Classification: 
1. Supports PTLDS:
   - Attributes persistent symptoms after successful Lyme disease treatment to other factors like 
   immune dysfunction, chronic fatigue syndrome, or depression.
   - Opposes the use of prolonged or repeated antibiotic treatments for PTLDS.
   - Suggests or implies that CLD lacks scientific support.

2. Supports CLD:
   - A contested condition wherein some believe symptoms result from an ongoing or active *Borrelia 
   burgdorferi* infection after standard treatment, often cited as requiring extended antibiotic therapy.
   - Suggests that persistent symptoms may require prolonged or repeated antibiotic treatment.

3. Neutral:
   - Presents balanced arguments without clearly supporting or refuting either PTLDS or CLD.
   - Discusses aspects of Lyme disease that relate to both sides of the PTLDS and CLD debate without 
   taking a clear stance.
   - No clear stance or support or refutation is provided regarding PTLDS or CLD, either explicitly or
   implicitly.

4. Unrelated:
   - The abstract does not mention PTLDS or CLD, or the surrounding debate at all.
   - Focuses on other aspects of Lyme disease, such as acute Lyme disease, vector control, ecological 
   studies, epidemiology, or diagnostic methods without mentioning chronic conditions.
   - References CLD without any explicit or implied previous treatment. 

5. Animal Study:
   - The abstract pertains exclusively to animal models or non-human subjects without direct implications
   for PTLDS or CLD in humans.
   - Does not involve human clinical data or direct conclusions about PTLDS or CLD in humans.
   - If the animal study provides evidence directly relevant to PTLDS or CLD in humans, classify it
   accordingly under "Supports PTLDS", "Supports CLD", or "Neutral".


Additionally, assign a Confidence Level  to each classification you make: High, Medium, or Low, based 
on how confident you are in the given classification.

- High: The chosen classification is indisputably true and cannot possibly be any of the other options.
The abstract contains explicit wording relating to its stance.
- Medium: The chosen classification requires a mix explicit and implicit evidence in the abstract. There
is some degree of certainty, but the classification is inferred rather than explicitly stated. Typically
arises from indirect language, ambiguous terms, or implied support for a stance.
- Low: The classification is uncertain due to lack of clear evidence or conflicting information in the 
abstract, relying more on inference and contextual understanding. Typically occurs when the abstract is
vague, uses ambiguous language, or includes evidence that could be interpreted in multiple ways. Reflects 
significant uncertainty in the classification, but that the chosen classification is more likely than the 
others. 
 
Also, provide a reason    comprising 2-3 sentences for your classification above. Clearly state your 
justification and if the abstract implicitly or explicitly supports PTLDS or CLD in humans. When determining
the abstract's stance, consider both explicit statements and implications of the abstracts given context 
and chosen topic. Look for language that suggests support or opposition even if not directly stated.
Input:
For each abstract, you will receive an index number of the abstract, a paper title, and the abstract text.

JSON Output Structure: 

Your output must be a JSON object containing the following fields:
- `index`: The index number of the abstract, which MUST match the input.
- `classification`: One of "Supports PTLDS", "Supports CLD", "Neutral", "Unrelated", or "Animal Study"
- `confidence`: Confidence level in the classification ("High", "Medium", or "Low").
- `reason`: Provide a 2-3 sentence justification for your classification.

< few-shot learning examples comprising 3 abstracts and their classification >

ABSTRACT:
< input >
\end{verbatim}
\end{mdframed}

\subsection{Example classification results post self-reflection}
\label{app:selfreflection}

A selection of abstracts whose classification changed from one category to the next post self-reflection prompting together with the justification texts, are shown in Table \ref{tab:example-classifications}.

\begin{table}[!htbp]
\centering
\fontsize{7pt}{9pt}\selectfont
\caption{Selected examples of abstracts with original and revised classifications and justifying text.}
\label{tab:example-classifications}
\begin{tabular}{p{7cm} p{0.80cm} p{0.80cm} p{3.5cm} p{3.5cm}}
\toprule
\textbf{Abstract} & \textbf{Original Classification} & \textbf{Revised Classification} & \textbf{Original Justification} & \textbf{Revised Justification}\\
\midrule

\textit{Background: IL-26 has demonstrated antimicrobial properties, as well as in the degradation of DNA from the Lyme disease spirochete Borrelia burgdorferi (Bb). Additionally, IL-26 can promote macrophage activation and enhance Bb phagocytotic activity. It is unclear if cell-mediated immune responses are modulated through TLR9 signaling when exposed to IL-26 Bb DNA complexes in post-treatment Lyme disease syndrome (PTLDS). Objective: We here aim to explore the effect of IL-26 in human Toll-like receptor (TLR)-9’s activation upon the recognition of Bb DNA. Methods: We utilized a single-receptor cell system, HEK-DualTM hTLR9 cells, which harbors two reporter plasmids for the NF-$κB$ and IL-8 signaling pathways. Bb DNA was exposed to increasing concentrations of IL-26 in monomeric or dimeric form and then used to stimulate the cells for 4 h. The TLR-9 ligand CpG was used as a control. Results: We observed that NF-$κ$B and IL-8 activation was maximal when the cells were stimulated with Bb DNA that had been treated with 5 $μM$ of IL-26 monomer and 1 $μM$ of IL-26 dimer. The same was observed for IL-8 activation upon CpG stimulation. We observed, however, a decrease in NF-$κB$ activation when treated with either form of IL-26. An NF-$κ$B activation increase did not occur with IL-26-treated TLR9 ligand CpG. Conclusions: Our study shows an enhancement in NF-$κ$B and IL-8 activation upon the recognition of IL-26-treated Bb DNA by TLR9, which suggests an increase in sensing by the TLR9 of Bb DNA when it is in the form of an IL-26-Bb DNA complex. These findings will prompt further studies on the interaction between IL-26 and Bb DNA. 145} 
& Neutral 
& Supports PTLDS 
& The abstract discusses the role of IL-26 in the immune response to Borrelia burgdorferi DNA and its potential implications for post-treatment Lyme disease syndrome (PTLDS). However, it does not take a clear stance on whether persistent symptoms are due to ongoing infection or other factors, nor does it advocate for or against the concepts of PTLDS or chronic Lyme disease (CLD). The focus is primarily on the biochemical interactions and immune mechanisms without explicit support for either side of the debate.
& Upon reevaluation, the abstract explicitly mentions 'post-treatment Lyme disease syndrome (PTLDS)' and explores immune responses related to this condition. The focus on IL-26's role in enhancing immune sensing of Bb DNA suggests a mechanism that aligns with the PTLDS perspective, which attributes persistent symptoms to immune dysfunction rather than ongoing infection. This indicates a clear inclination towards supporting PTLDS, justifying a revision to 'Supports PTLDS' with a High confidence level.\\
\midrule

\textit{Arthritis is the most common late manifestation of Borrelia burgdorferi infection in the United States, usually beginning months after the tick bite. In most patients with Lyme arthritis (LA) today, arthritis is the presenting manifestation of the disease. Patients have swelling and pain in one or a few large joints, especially the knee. Serologic testing is the mainstay of diagnosis. Responses to antibiotic treatment are generally excellent, although a small percentage of patients have persistent, postinfectious synovitis after 2 to 3 months of oral and IV antibiotics, which respond to anti-inflammatory therapies. Herein we review the clinical presentation, diagnosis, and management of LA. 7787}
& Unrelated
& Neutral
& The abstract focuses on Lyme arthritis as a manifestation of Borrelia burgdorferi infection, discussing its clinical presentation, diagnosis, and management. While it mentions persistent postinfectious synovitis, it does not engage with the debates surrounding PTLDS or CLD, nor does it imply a stance on these conditions. The content is primarily concerned with the acute and post-treatment aspects of Lyme arthritis rather than chronic conditions.
& Upon reevaluation, the abstract does mention persistent postinfectious synovitis, which could be interpreted as a connection to ongoing symptoms after treatment, potentially linking it to the PTLDS debate. However, it does not explicitly support or refute either PTLDS or CLD, making it more appropriate to classify it as Neutral rather than Unrelated. The confidence level is adjusted to Medium, as the abstract does touch on relevant themes but lacks a clear stance.\\
\midrule

\textit{The immuno-pathogenetic mechanisms underlying chronic Lyme neuroborreliosis are mainly unknown. Human Borrelia burgdorferi (Bb) infection is associated with Bb-specific secretion of interferon-γ (IFN-γ), which may be important for the elimination of Bb, but this may also cause tissue injury. In order to increase the understanding of the pathogenic mechanisms in chronic neuroborreliosis, we investigated which cell types that secrete IFN-γ. Blood mononuclear cells from 13 patients with neuroborreliosis and/or acrodermatitis chronicum atrophicans were stimulated with Bb antigen and the phenotypes of the induced IFN-γ-secreting cells were analyzed with three different approaches. Cells expressing CD8 or TCRγδ, which both have cytolytic properties, were the main phenotypes of IFN-γ-secreting cells, indicating that tissue injury in chronic neuroborreliosis may be mediated by cytotoxic cells.}
& Supports CLD
& Neutral
& The abstract discusses the immunological response in patients with chronic Lyme neuroborreliosis, suggesting that the secretion of interferon-γ by specific cell types may contribute to tissue injury. This aligns with the CLD perspective, which posits that ongoing immune responses and potential persistent infection could lead to chronic symptoms. However, the abstract does not explicitly advocate for prolonged antibiotic treatment or directly state that ongoing infection is the cause of symptoms, leading to a medium confidence classification.
& Upon reevaluation, the abstract primarily focuses on the immunological mechanisms involved in chronic Lyme neuroborreliosis without explicitly supporting the notion of ongoing infection or the need for prolonged antibiotic treatment. While it discusses the role of immune responses in tissue injury, it does not provide clear evidence that aligns with the CLD perspective. Therefore, it is more accurately classified as Neutral, as it does not take a definitive stance on either PTLDS or CLD.\\

\bottomrule
\end{tabular}
\end{table}


\section{Inter-rater Reliability Web Application}
\label{appendix:WebApplication}

\subsection{Web App with Example Texts and Outputs}

\begin{figure}[hbtp]
    \centering
    \framebox[\textwidth] {\includegraphics[width=\linewidth]{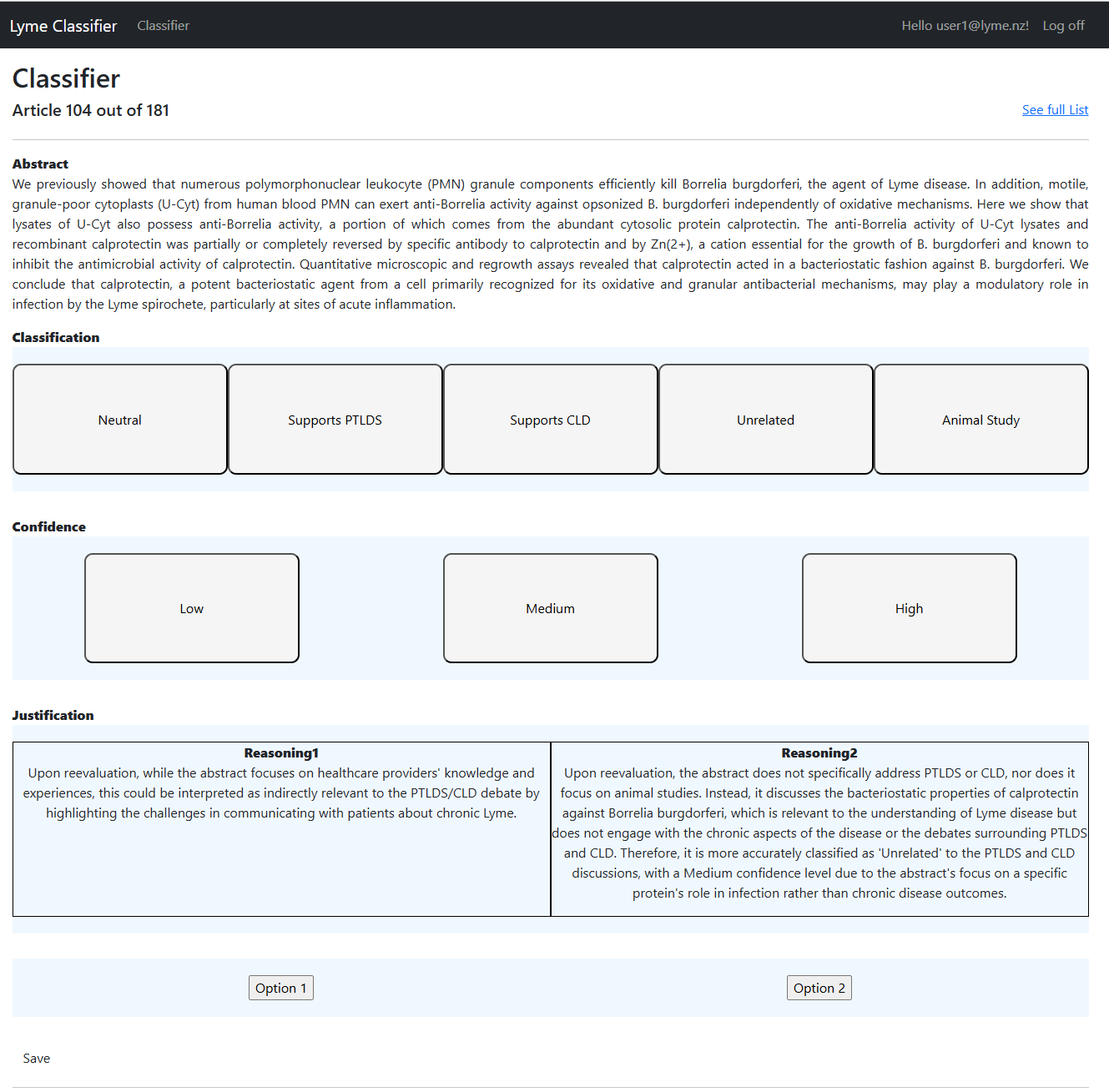}}
    \caption{IRR web app showing the user abstract classification interface}
    \label{fig:web_app_example}
\end{figure}

Figure \ref{fig:web_app_example} illustrates our custom-built web application created specifically for the human validation process (Step 3 in Figure \ref{fig:meth}). In this interface, a user (in this case, a subject-expert rater) logs in to evaluate a given abstract. The application displays the original abstract text and prompts the user to choose one of several classification labels (e.g., Supports PTLDS, Supports CLD, Neutral, or Animal Study). Next, the rater selects a confidence level in their choice (e.g., Low, Medium, or High) and selects one of the two options that best describes the justification for the classification.

\begin{figure}[hbtp]
    \centering
    \framebox[\textwidth] {\includegraphics[width=\linewidth]{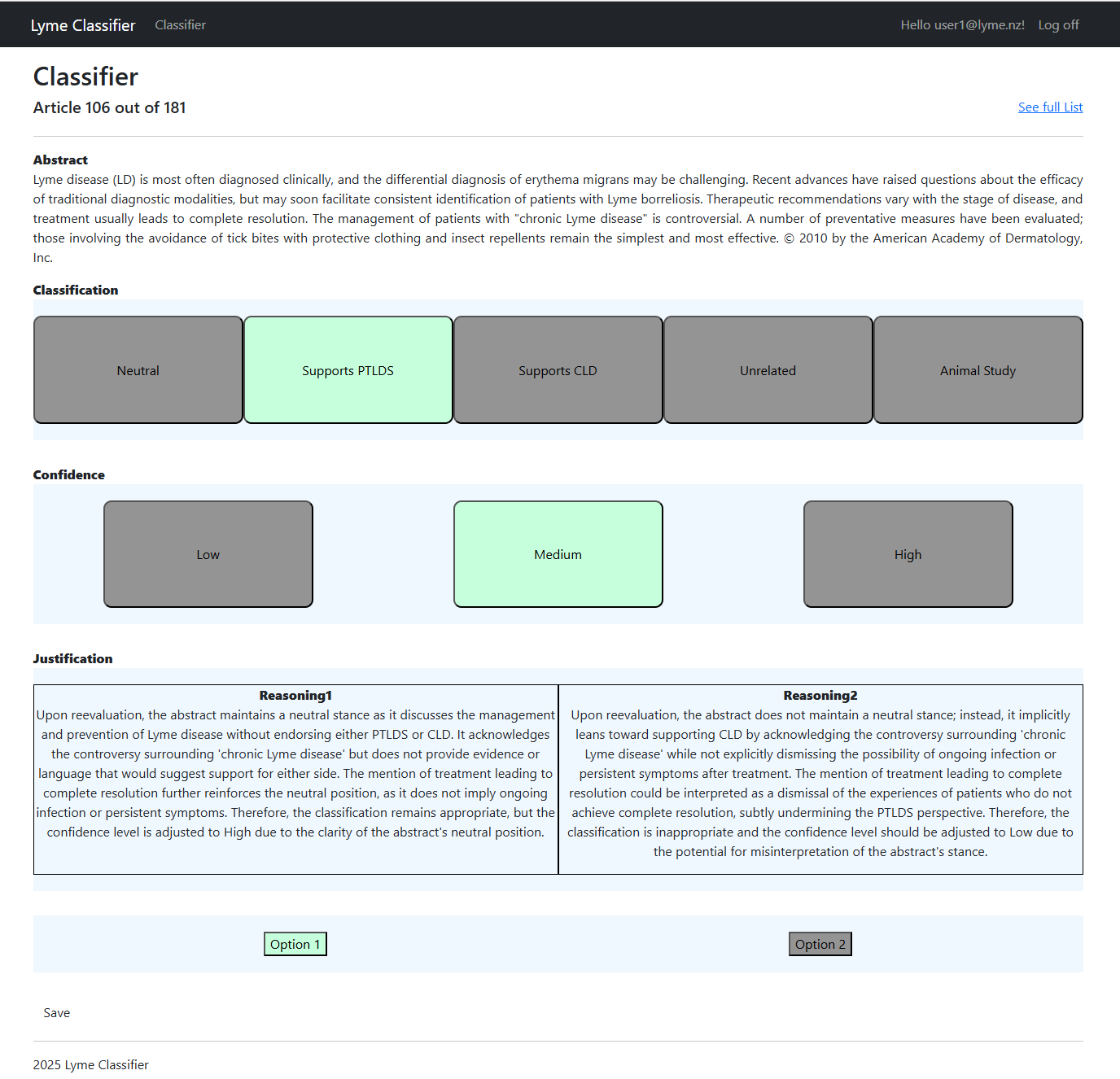}}
    \caption{IRR web app demonstrating the classification process}
    \label{fig:web_app_demonstration}
\end{figure}

Figure \ref{fig:web_app_demonstration}  depicts another view of the same web application but for a different abstract. This example emphasizes the classification process itself, highlighting how a user can read through the abstract, reflect on the core content (e.g., whether it focuses on CLD, PTLDS, or otherwise), and then decide upon the appropriate classification label. The interface displays the choices that a user has selected in green.
Together, both figures demonstrate the workflow that expert human raters followed in the IRR analysis process to validate the AI-based stance detection and justification texts provided for the rationale behind the LLM's decisions.

\end{document}